\documentclass[aps,  showpacs]{revtex4}
\usepackage{graphicx}

\begin{document}
\title{Nonlinear Response Functions of Strongly Correlated Boson Fields: Bose-Einstein Condensates
 and Fractional Quantum Hall Systems}
\author{Stephen Choi$^{1}$, Oleg Berman$^{2}$,  Vladimir Chernyak$^{3}$, and Shaul
Mukamel$^{1,2}$} \affiliation{\mbox{$^{1}$ Department of Physics
and Astronomy, University of Rochester,
 Box 270216,  Rochester,} \\ New York 14627-0216 \\
\mbox{$^{2}$ Department of Chemistry, University of Rochester, Box 270216,  Rochester,} \\ New York 14627-0216 \\
 \mbox{$^{3}$ Corning Incorporated,
Process Engineering and Modeling, Corning, New  York 14831} }

\date{\today}

\vspace{0.5cm} \preprint{{\em Submitted to Phys. Rev. {\bf A} }
\hspace{3.7in} Web galley}

\vspace{3cm}

\begin{abstract}
The second order response functions and susceptibilities of finite
temperature Bose-Einstein Condensates (BEC) in a one dimensional
harmonic trap driven by an external  field that couples to the
particle density are calculated by solving the time-dependent
Hartree-Fock-Bogoliubov (TDHFB) equations. These provide
additional insight into BEC dynamics, beyond the linear response
regime. The results also apply to electron liquids in the
Fractional Quantum Hall Effect (FQHE) regime which can be mapped
onto an effective boson system coupled to a Chern-Simons gauge
field.
\end{abstract}

\pacs{03.75.Fi}

\maketitle

\section{Introduction}

We have recently studied an externally driven, finite temperature
Bose-Einstein condensates (BEC) described by the time-dependent
Hartree-Fock-Bogoliubov (TDHFB) equations~\cite{choi}. A
systematic procedure was outlined for solving these equations
perturbatively  in the applied external field, and
position-dependent linear response functions and susceptibilities
were calculated.

In this paper we extend this formalism to calculate the second
order response function for the condensate, non-condensate density
and non-condensate correlation. The linear response provides an
adequate description of the system only for weak external
perturbations. Otherwise, nonlinear effects contained in the
higher order terms in the perturbation series may not be ignored.
There are strong similarities between nonlinear optics and the
dynamics of BEC owing to the interatomic interactions in Bose
condensates. Previous work done in this area includes the
demonstration of four wave mixing in zero temperature BEC using
the Gross-Pitaevskii Equation (GPE)~\cite{Band}. In the present
article we  study nonlinear properties of BEC using the  TDHFB
framework.

Numerous dynamical theories exist for finite temperature BEC that
takes into account higher order collision processes, such as the
time-dependent Bogoliubov-de Gennes equations~\cite{CastinDum},
the Hartree-Fock-Bogoliubov (HFB)
theory~\cite{Ring,Blaizot_Ripka,Griffin,Prou2}, Quantum Kinetic
Theory~\cite{Baym,Martin,Gardiner,ZNG,Walser,Zaremba}, and
Stochastic
methods~\cite{Krauth,Ceperley,Walls,Drummond,Carusotto}. The TDHFB
theory is a self-consistent theory of BEC in the collisionless
regime that progresses logically from the Gross-Pitaevskii
Equation by taking into account higher order correlations of
noncondensate operators. Although TDHFB neglects higher order
correlations included in the various quantum kinetic theories, the
TDHFB equations are valid at very low temperatures near zero, even
down to the zero temperature limit, and are far simpler than the
kinetic equations which can only  be solved using approximations
such as ZNG. Another attractive feature of TDHFB from a  purely
pragmatic point of view is that the Fermionic version of the
theory has already been well-developed in Nuclear
Physics~\cite{Ring}. We therefore work at the TDHFB level  in this
paper and our approach draws upon the analogy with the
time-dependent Hartree-Fock (TDHF) formalism developed for
nonlinear optical response of many electron systems~\cite{TDHF}.

In Section II we introduce the second order time and frequency
domain response functions for an externally driven BEC. Numerical
results  are discussed in Sections III and IV for a condensate of
2000 atoms in a one dimensional harmonic trap. In Section V we
show how this formalism may be applied for computing the second
order response for a Fractional Quantum Hall system. In Section VI
we conclude. Details of the derivation of the second order
response functions are given in the Appendix.

\section{Nonlinear response function of externally driven BEC}

We adopt the notation of Ref.~\cite{choi} throughout the paper.
The Hamiltonian describing the system of an externally driven,
trapped atomic BEC is given by:
\begin{equation}
\label{hamb} \hat{H} = \hat{H}_0 + \hat{H}^{\prime}(t) ,
\end{equation}
where
\begin{equation}
\hat{H}_0 = \sum_{ij} H^{sp}_{ij}\hat{a}^{\dagger}_{i}\hat{a}_{j} +
\frac{1}{
2}\sum_{ijkl} V_{ijkl} \hat{a}^{\dagger}_{i}\hat{a
}^{\dagger}_{j}\hat{a}_{k}\hat{a}_{l},  \label{Hamiltonian}
\end{equation}
and
\begin{equation}
H^{\prime}(t) = \eta \sum_{ij} E_{ij}(t) \hat{a}^{\dagger}_{i}\hat{a}_{j}
\end{equation}
with
\begin{equation}
E_{ij}(t)  \equiv \int \! d^{3}{\bf r} \, \phi^{*}_{i}({\bf r})
V_{f}({\bf r},t) \phi_{j}({\bf r})
\hat{a}^{\dagger}_{i}\hat{a}_{j}. \label{E_t} \label{hambo}
\end{equation}
The boson operators $\hat{a}^{\dagger}_{i}$
( $\hat{a}_{i}$ ) create ( annihilate ) a particle from
a basis state with wave functions $\phi_{i}({\bf r})$.
The single particle Hamiltonian $H^{sp}$ in Eq. (\ref{Hamiltonian}) is diagonal if the basis state $\phi_{i}({\bf
r})$  is chosen to be the eigenstates of the trap, while the symmetrized interaction
matrix elements,
\begin{equation}
V_{ijkl} = \int \! d^{3}
{\bf r} \, d^{3}{\bf r^{\prime}} \, \phi^{*}_{i}({\bf r})\phi^{*}_{j}({\bf
r^{\prime}})V({\bf r}-{\bf r^{\prime}})\phi_{k}({\bf
r^{\prime}})\phi_{l}({\bf
r}) , \label{Vijkl}
\end{equation}
describe the collision between the atoms, with $V({\bf r}-{\bf
r^{\prime}})$ being a general interatomic potential.
$H^{\prime}(t)$ describes the effect of a general external force
$V_{f}({\bf r},t)$ on the condensate that mimics the mechanical
force applied experimentally such as shaking of the
trap~\cite{ExRb,ExNa}.

The dynamics of the system is calculated by solving the
time-dependent Hartree-Fock-Bogoliubov (TDHFB) equations for the
condensate mean field, $z_{i} =  \langle \hat{a}_{i} \rangle$, the
non- condensate density $\rho_{ij} = \langle \hat{a}_{i}^{\dagger}
\hat{a}_{j} \rangle - \langle \hat{a}_{i}^{\dagger} \rangle
\langle \hat{a}_{j} \rangle$, and the non-condensate correlations
$\kappa_{ij} =  \langle \hat{a}_{i} \hat{a}_{j} \rangle - \langle
\hat{a}_{i} \rangle \langle \hat{a}_{j} \rangle$. These are
presented in Appendix \ref{TDHFBEqs}. These nonlinear coupled
equations are solved by an order by order expansion of the
variables $z_{i}$,  $\rho_{ij}$ and  $\kappa_{ij}$; at each order,
the resulting equations to be solved  become linear~\cite{choi}.
It is found that the sequence of linear equations to be solved has
the general form:
\begin{equation}
i\hbar \frac{d\vec{\psi}^{(n)}(t)}{dt} = {\cal L}^{(n)}
\vec{\psi}^{(n)}(t) + \lambda^{(n)}(t) , \label{general}
\end{equation}
where we have denoted the set of $n$'th order variables
$z^{(n)}_{i}$, $\rho^{(n)}_{ij}$, and $\kappa^{(n)}_{ij}$  as a
$2N(2N+1) \times 1$ column vector in the Liouville space notation
where $N$ is the number of basis states used, $\vec{\psi}^{(n)} =
[\vec{z}^{(n)}, \vec{z}^{(n)*},
\vec{\rho}^{(n)},\vec{\kappa}^{(n)},\vec{\rho}^{(n)*},
\vec{\kappa}^{(n)*}]^{T}$, and the $2N(2N+1) \times 2N(2N+1)$
matrix ${\cal L}^{(n)}$ is the $n$-th order Liouville operator
obtained from the TDHFB equations~\cite{choi}.  The matrices
${\cal L}^{(n)}$ for all orders $n > 0$ are identical i.e. ${\cal
L}^{(1)} \equiv {\cal L}^{(2)} \equiv  \cdots {\cal L}^{(n)}$ so
that only the matrices ${\cal L}^{(0)}$ and ${\cal L}^{(1)}$ are
required to be calculated. ${\cal L}^{(0)}$ and ${\cal L}^{(1)}$
are presented in Appendix \ref{Lmatrix}. The formal solution to
Eq. (\ref{general}) is:
\begin{equation}
\vec{\psi}^{(n)}(t) = \frac{1}{i\hbar }\int_{0}^{t}\exp \left[ -\frac{i}{
\hbar
}{\cal L}^{(n)}(t-t^{\prime })\right] \lambda^{(n)} (t^{\prime }) dt^{\prime
}.
\end{equation}
In the frequency domain, the solution to Eq. (\ref{general}) takes the form:
\begin{equation}
\vec{\psi}^{(n)}(\omega) = \frac{1}{\omega - {\cal L}^{(n)}}
\lambda^{(n)}(\omega),
\end{equation}
where $\vec{\psi}^{(n)}(\omega)$ and $\lambda^{(n)}(\omega)$ are the Fourier
transforms  of $\vec{\psi}^{(n)}(t)$ and  $\lambda^{(n)}(t)$ respectively.

For the zero'th order ($n=0$), we obtain the time- independent HFB
equations (TIHFB)
\begin{equation}
{\cal L}^{(0)} \vec{\psi}^{(0)}(t) = 0 ,
\end{equation}
while for the first order ($n=1$),  the equation solved is Eq.
(\ref{general}) with  $\lambda^{(1)}(t)$ being a $2N(2N+1) \times
1$ vector $\zeta(t)$ calculated in Ref.~\cite{choi}; $\zeta(t)$ is
also presented in Appendix \ref{Lmatrix}.

Once  the $n$'th order solution to TDHFB is found, we can proceed
to define the $n$'th order response functions. The physical
significance of the response functions become more transparent
when the  $n$'th order solutions $\alpha^{(n)}$  where $\alpha$ is
one of the variables $z$, $\rho$ or $\kappa$  are expressed in
real space.  We therefore introduce the position dependent
variables written in terms of the trap eigenstate basis:
\begin{eqnarray}
z^{(n)} ({\bf r}, t) & = & \sum_{j} z^{(n)}_{j}(t) \phi_{j} ({\bf r}), \label{z_r} \\
\rho^{(n)} ({\bf r}, t) & = & \sum_{ij} \rho^{(n)}_{ij}(t) \phi^{*}_{i} ({\bf r})\phi_{j} ({\bf r}),   \label{rho_r} \\
\kappa^{(n)} ({\bf r}, t) & = & \sum_{ij} \kappa^{(n)}_{ij}(t) \phi_{i} ({\bf r}) \phi_{j} ({\bf r}). \label{kappa_r}
\end{eqnarray}

Real space non-condensate density and non-condensate correlations
are, in general, nonlocal functions of two spatial points
$\rho({\bf r}', {\bf r})$ and $\kappa({\bf r}', {\bf r})$. We only
computed these quantities for ${\bf r} = {\bf r}'$ in this paper
since these are the most physically accessible.  Measuring these
quantities with ${\bf r} \neq {\bf r}'$ involves observing atomic
correlations which is much more difficult than  photon
correlations. In Liouville space notation, the position-dependent
$n$th order solution  $\vec{\psi}^{(n)}({\bf r}, t)$ can be
defined using the relations Eq. (\ref{z_r} - \ref{kappa_r}) and
introducing a $2N(2N+1) \times 2N(2N+1)$ square matrix
$\tilde{\Upsilon}({\bf r})$:
\begin{equation}
\vec{\psi} ^{(n)} ({\bf r}, t) \equiv \tilde{\Upsilon}({\bf r}) \vec{\psi}^{(n)} (t)
\end{equation}
where
\begin{equation}
\tilde{\Upsilon}({\bf r}) = {\rm diag} \left [ \tilde{\phi}({\bf r}),
\tilde{\phi}^{*}({\bf r}), \Phi_{\rho}({\bf r}), \Phi_{\kappa}({\bf r}),
\Phi_{\rho}^{*}({\bf r}), \Phi_{\kappa}^{*}({\bf r}) \right ] . \label{upsilontilde}
\end{equation}
Here ``${\rm diag}[ \cdots ]$'' denotes that
$\tilde{\Upsilon}({\bf r})$ is  a block diagonal square matrix
made of $N \times N$ blocks $\tilde{\phi}({\bf r})$,
$\tilde{\phi}^{*}({\bf r})$ and $N^2 \times N^2$ blocks
$\Phi_{\rho}({\bf r}), \Phi_{\kappa}({\bf r}),
\Phi_{\rho}^{*}({\bf r}), \Phi_{\kappa}^{*}({\bf r})$.
$\tilde{\phi}({\bf r})$ is a diagonal matrix with the $i$-th
diagonal element given by the basis states $\phi_{i}({\bf r})$,
and  $\Phi_{\rho}({\bf r})$ and $\Phi_{\kappa}({\bf r})$ are also
diagonal matrices whose $ij$'th diagonal element  are given by
$\left [ \Phi_{\rho}({\bf r}) \right ]_{ij,ij} = \phi^{*}_{i}({\bf
r})\phi_{j}({\bf r})$, and $\left [ \Phi_{\kappa}({\bf r}) \right
]_{ij,ij} = \phi_{i}({\bf r})\phi_{j}({\bf r})$ respectively. The
real space variables $z^{(n)} ({\bf r}, t)$,  $\rho^{(n)} ({\bf
r}, t)$, and $\kappa^{(n)} ({\bf r}, t)$ are finally obtained by
summing over the appropriate elements of the vector
$\vec{\psi}^{(n)} ({\bf r}, t)$:
\begin{equation}
z^{(n)} ({\bf r}, t) = \sum_{i=1}^{n}  \vec{\psi}_{i}^{(n)} ({\bf
r}, t) , \;\;\;\; \rho^{(n)} ({\bf r}, t) = \sum_{i =
2n+1}^{2n+n^2}  \vec{\psi}_{i}^{(n)} ({\bf r}, t) , \;\;\;\;
\kappa^{(n)} ({\bf r}, t) = \sum_{i = 2n + n^2+1}^{2n + 2n^2}
\vec{\psi}_{i}^{(n)} ({\bf r}, t) .
\end{equation}

The position-dependent second order ($n=2$) response function
$K^{(2)}_{\alpha}(t,t_1,  t_2, {\bf r},{\bf r}_1, {\bf r}_2)$ is
then defined as follows:
\begin{equation}
\alpha^{(2)}({\bf r},t) = \int K^{(2)}_{\alpha}(t,t_1, t_2,
{\bf r},{\bf r}_1, {\bf r}_2), V_{f}({\bf r}_1,t_1) V_{f}({\bf r}_2,t_2)
dt_1
dt_{2}  d{\bf r}_1  d {\bf r}_2,
\end{equation}
where $\alpha^{(2)}({\bf r},t)$ are second order
solutions,
$\alpha = z, \rho, \kappa$. Expression for $K^{(2)}$ is given in Appendix
\ref{K2}, Eqs. (\ref{K2tz})-(\ref{K2tkappa}).
Having found the time domain response $K^{(2)}_{\alpha}(t,t_1, t_2,
{\bf r},{\bf r}_1, {\bf r}_2)$, the second order susceptibility is obtained
by a Fourier transform to the frequency domain:
\begin{equation}
\chi^{(2)}_{\alpha}(\Omega, \Omega_{1}, \Omega_{2}, {\bf r},{\bf r}_1, {\bf
r}_2)   =
\int_{0}^{\infty} dt \int_{0}^{\infty} dt_{1} \int_{0}^{\infty}
dt_{2} K^{(2)}_{\alpha}(t, t_{1}, t_{2}, {\bf r},{\bf r}_1, {\bf r}_2) \exp
\left (i \Omega t + i \Omega_{1} t_{1} + i \Omega_{2} t_{2} \right ).
\end{equation}
A closed expression for $\chi^{(2)}$ used in our numerical
calculations is given in Appendix \ref{chi2}, Eqs.
(\ref{K2wz})-(\ref{K2wkappa}).

\section{Time domain response}

So far, all our results were given in the trap basis, and hold for
a general interatomic interaction potential.  In the following
numerical calculations, we approximate the interatomic potential
$V({\bf r}-{\bf r^{\prime}})$ in Eq.~(\ref{Vijkl}) by a contact
potential:
\begin{equation}
\begin{array}{c@{\hspace{1cm}}c}
{\displaystyle V({\bf r}-{\bf r^{\prime}}) \rightarrow U_{0} \delta({\bf
r}-%
{\bf r^{\prime}}), } & {\displaystyle U_{0} = \frac{4\pi\hbar^{2}a}{m}, }
\end{array}
\label{contactpotential}
\end{equation}
where $a$ is the {\em s}-wave scattering length and $m$ is the atomic mass.
This is valid because wave functions at ultracold temperatures have very long
wavelengths compared to the range of interatomic potential implying that
details of the interatomic potential become unimportant.
The tetradic matrices $V_{ijkl}$ defined in Eq. (\ref{Vijkl}) are then
simply given by:
\begin{equation}
V_{ijkl} = \frac{4\pi\hbar^{2}a}{m} \int \phi^{*}_{i}({\bf r})\phi^{*}_{j}(%
{\bf r})\phi_{k}({\bf r})\phi_{l}({\bf r}) d{\bf r}.
\end{equation}

We consider a 2000 atom one dimensional condensate in a harmonic
trap. The parameters used for our numerical calculation of
$\vec{\psi}^{(0)}$ are: $U_0 = \frac{4\pi\hbar^{2}a}{m} = 0.01$,
and  temperatures $0 \hbar \omega_{\rm trap}/k$ and $10 \hbar
\omega_{\rm trap}/k$ where $\omega_{\rm trap}$ is the trap
frequency, $k$ is the Boltzmann constant and the basis set size of
$N=5$ is used.  We keep the trap units throughout with 256 grid
points for position.  The same parameters were used in the
calculations of the linear response in Ref.~\cite{choi}.

To solve for the second order response, both the zero'th and the
first order solutions must be found. Calculation of the zero'th
order solution from the TIHFB equations is the most numerically
involved step, as it requires solving nonlinear coupled equations.
Griffin has provided a self-consistent prescription for solving
the TIHFB, in terms of the Bogoliubov-de Gennes
Equations~\cite{Griffin}. We have therefore followed the
prescription of Ref.~\cite{Griffin} to find the solution to TIHFB.
Once the zero'th order solution is found, it is straightforward to
calculate the first and the second order response functions. The
calculation of the eigenvalues of the non-Hermitian matrix ${\cal
L}^{(2)}$ required for computing the response functions was
carried out using the Arnoldi algorithm~\cite{Arnoldi}.

In order to provide an indication of the structure of the matrix
${\cal L}^{(2)}$, we first plot in Fig.~1 the {\it linear}
susceptibility $K^{(1)}(\Omega, {\bf r}=0, {\bf r}_1=0)$  for zero
and finite temperatures. Peak positions indicate the resonant
frequencies.

An expression for the second order time domain response function
is given in Appendix \ref{K2}, Eqs. (\ref{K2tz}) -
(\ref{K2tkappa}). We have first obtained the numerical solution to
TIHFB, $2N(2N+1) \times 1$ vector $\vec{\psi}^{(0)}$ evaluated at
zero and finite temperatures, the $2N(2N+1) \times 2N(2N+1)$
matrices $\tilde{\Upsilon}$, ${\cal U}$, and $\tilde{\Phi}$
defined in Eqs. (\ref{upsilontilde}), (\ref{U}), and
(\ref{phitilde}), and the $2N(2N+1) \times 1$ vector $\vec{\Xi}_K$
defined in Eqs. (\ref{XiK}-\ref{XiKkappa}).  Substituting these
into Eqs. (\ref{KIt1}-\ref{KIIt1}), the final calculation involves
matrix multiplication of  $2N(2N+1) \times 1$ vectors
$\vec{\psi}^{(0)}$ and $\vec{\Xi}_K$ with $2N(2N+1) \times
2N(2N+1)$ matrices $\tilde{\Upsilon}$, ${\cal U}$, and
$\tilde{\Phi}$, and integration over the time variable $\tau$.
$\tilde{\Upsilon}$ and $\tilde{\Phi}$ are constructed in terms of
the harmonic oscillator basis states which are calculated
numerically from the recursive formula that involves the Gaussian
function multiplying the Hermite polynomials~\cite{Arfken}. The
matrix ${\cal U}$ is calculated using a MATLAB function that uses
the Pad\'{e} approximation for matrix exponentiation~\cite{Golub}.

We present the second-order response function in the time domain
$K^{(2)}(t,t_1,t_2, {\bf r}, {\bf r}_1 {\bf r}_2)$ as a function
of ${\bf r}$ and ${\bf r}_1$ at various times $t,t_1,t_2$ and
${\bf r}_2$. This provides a way to depict graphically the
correlation involving 6 variables $t$, $t_1$, $t_2$,  ${\bf r}$,
${\bf r}_1$,and ${\bf r}_2$ on a 2 dimensional plot, and gives a
``snapshot'' of the position-dependent second order correlations
across the condensate. The times $t$,  $t_1$, $t_2$, are
respectively the time of detection and the time of the first and
second applied short fields, while ${\bf r}$, ${\bf r}_1$,and
${\bf r}_2$ denote the corresponding spatial variables. The
position-dependence is important since the experimentally produced
condensates are mesoscopic in size; in optical spectroscopy,
however, the dipole approximation usually applies and consequently
the spatial dependence of the response is irrelevant.  Fig. 2
shows the absolute value of the second order response functions in
the time domain with the time $t_2$ fixed at $t_2 = 0$.  Fig. 2
(a) is for the position of perturbation fixed at ${\bf r}_2 = 0$
i.e. the center of the trapped atomic cloud while Fig. 2 (b) is
for ${\bf r}_2 = -5$ at the edge of the cloud. All positions are
referred to in  harmonic oscillator length units.   The plots are
for zero temperature condensate at the short, intermediate and
long times ($\{t, t_1  \} = \{5.89, 2.6 \}$, $\{t, t_1 \} =
\{15.7,  7.2 \}$, $\{t, t_1 \} = \{ 31.4, 15.7 \}$) as indicated
at the top of each column of figures. The times are given in units
of the trap period, $1/\omega_{\rm trap}$. The top, middle, and
bottom rows give the response function for the condensate $z$,
non-condensate density $\rho$, and non-condensate correlation
$\kappa$ respectively. The dashed circle represents the spatial
extent of the trapped BEC. The corresponding plot for a finite
temperature BEC at temperature $T = 10 \hbar \omega_{\rm trap} /k$
is displayed in Fig. 3.

The spatially asymmetric functions at ${\bf r}_2 = -5$ reverses
its shape for ${\bf r}_2 = 5$, with similar changes also observed
for the ${\bf r}_2 = \pm 2.5$ pair. With ${\bf r}_2 = \pm 2.5$
which is a point inside the atomic cloud, the contours had
markedly less symmetric shape than for ${\bf r}_2 = \pm 5$ at the
very edge of the atomic cloud. The response functions take a more
symmetric shape as time increases. The ${\bf r}_2$ dependence of
the response function is found to be reduced at longer times.
Comparing Figs. 2 and 3, the response functions clearly show
strong temperature dependence with the functions giving distinct
contours at different temperatures.  The response functions attain
spatial symmetry more rapidly at zero temperature owing to the
weaker coupling between the variables $z$, $\rho$ and $\kappa$.

\section{Frequency domain response}

Using Eq. (\ref{freqdom_K2}) we have calculated the second order
susceptibility.  Eq. (\ref{freqdom_K2}) is also a matrix
multiplication involving $2N(2N+1) \times 1$ vectors
$\vec{\psi}^{(0)}$ and $\vec{\Xi}_{K}(\omega)$ defined in Appendix
\ref{chi2}, Eq.~(\ref{XiKwkappa}), with the $2N(2N+1) \times
2N(2N+1)$ matrices $\tilde{\Upsilon}$, ${\cal U}(\omega)$, and
$\tilde{\Phi}$. The Green's function ${\cal U}(\omega)$ is
calculated as follows:
\begin{equation}
{\cal U}(\omega) = \frac{1}{\omega - {\cal L}^{(2)} + i \epsilon} =
\sum_{\nu}
\frac{\xi_{\nu}\zeta_{\nu}^{\dagger}}{\omega - \omega_{\nu} + i \epsilon} ,
\end{equation}
where $\xi_{\nu}$ is the right eigenvector of ${\cal L}^{(2)}$
with eigenvalues $\omega_{\nu}$ such that ${\cal L}^{(2)}
\xi_{\nu}  =  \omega_{\nu} \xi_{\nu}$ and $\zeta_{\nu}$ are the
left eigenvectors of ${\cal L}^{(2)}$ such that $\sum_{\nu}
\xi_{\nu}\zeta_{\nu}^{\dagger} =  \openone$. The eigenvalues
$\omega_{\nu}$ of ${\cal L}^{(2)}$ were calculated using the
Arnoldi algorithm~\cite{Arnoldi}.

The absolute value of the second order response function in the
frequency domain, $|K^{(2)}(\Omega,\Omega_1,\Omega_2, {\bf r},
{\bf r}_1 {\bf r}_2)|$ is displayed In Fig. 4 with the position
variable ${\bf r}_2$  set at (a) ${\bf r}_2 = 0$ at the center of
the atomic cloud, and (b) ${\bf r}_2 = -5$ at the edge of the
atomic cloud. The plots are for zero temperature condensate at
various frequencies $\Omega_1$ and $\Omega_2$ indicated at the top
of each column. We chose the frequencies such that $\Omega_1$,
$\Omega_2$, and $\Omega_1 + \Omega_2$ are off-resonant with
respect to the eigenvalues of ${\cal L}^{(2)}$ (first column,
$\Omega_1 = 2.23$, $\Omega_2 = 1.55$); both $\Omega_1$ and
$\Omega_2$ are on-resonance while $\Omega_1 + \Omega_2$ is
off-resonant (second column, $\Omega_1 = 2.2$, $\Omega_2 = 1.5$);
and finally $\Omega_1 + \Omega_2$ and $\Omega_2$ on-resonance
($\Omega_1 = 0.7$, $\Omega_2 = 1.5$). As with the previous
figures, the top, middle, and bottom rows give the response
function for the condensate, non-condensate density, and
non-condensate correlation respectively.  The result for a finite
temperature BEC at temperature $T = 10 \hbar \omega_{\rm trap} /k$
is displayed in Fig. 5. At finite temperature, the resonant
frequencies are shifted from the zero temperature counterpart so
that the actual frequency combinations used are the following:
$\Omega_1$, $\Omega_2$, and $\Omega_1 + \Omega_2$ off-resonant
($\Omega_1 = 2.45$, $\Omega_2 = 1.6$); both $\Omega_1$ and
$\Omega_2$ are on-resonance while $\Omega_1 + \Omega_2$ is
off-resonant ($\Omega_1 = 2.43$, $\Omega_2 = 1.5$); and
frequencies with $\Omega_1 + \Omega_2$ is on-resonance ($\Omega_1
= 0.92$, $\Omega_2 = 1.5$).

The susceptibilities become more spatially symmetric as resonant
frequencies are matched. The most symmetric function was generated
when the sum of two frequencies $\Omega_1$ and $\Omega_2$ was on
resonance while the least symmetric function resulted when both
frequencies were off-resonant. With only of one of the frequencies
on resonance, a result with an intermediate level of symmetry was
observed. The dependence of the response function on ${\bf r}_2$
is strongest for the off-resonant case, while the one with
$\Omega_1 + \Omega_2$ on resonance remains more or less unaffected
by the changes. As in the time domain response, the asymmetric
functions at ${\bf r}_2 = -5$ reverses its shape for ${\bf r}_2 =
5$. Such change was also observed for the ${\bf r}_2 = \pm 2.5$
pair; again, for ${\bf r}_2 = \pm 2.5$ which is inside the atomic
cloud the contours were found to be less symmetric than the
corresponding figures for ${\bf r}_2 = \pm 5$.

In Fig. 6 we plot the second order susceptibility at zero
temperature as a function of $\Omega_1$ and $\Omega_2$ with the
position variables set at ${\bf r} = {\bf r}_1 = {\bf r}_2 = 0$
i.e. at the center of the atomic cloud. The left column shows
$|K^{(2)}(\Omega_1,\Omega_2)|$ i.e. the absolute value of the
second order response function. The plot shows only few peaks near
the frequency of $1$ because the difference between the magnitude
of these highest few peaks and  the rest of the peaks occurring at
other frequencies is too large. This suggests that one should
ideally tune into the combination of frequencies $\Omega_1$ and
$\Omega_2$  corresponding to these highest peaks to observe the
maximum second order response experimentally. The second frequency
$\Omega_2$ implies a completely different physics compared to the
linear response; $\Omega_2$ is the frequency of a new harmonic
being generated as a result of a strong external perturbation
oscillating at frequency $\Omega_1$. The middle column of Fig. 6
gives  $|K^{(2)}(\Omega_1,\Omega_2)|$ where the largest peaks are
scaled down to the magnitude of the smaller peaks present. It
clearly shows a number of peaks present at frequencies $\Omega_1$
and $\Omega_2$ less than 1, and also  shows that with $\Omega_1$
($\Omega_2$) fixed at 1, there is a pronounced response  for many
values of $\Omega_2$ ($\Omega_1$). The right column  of Fig. 6
shows the logarithm of the left column.  This enables the large
variations in the magnitude of $|K^{(2)}(\Omega_1,\Omega_2)|$ to
be displayed. The top, middle, and bottom rows give the response
function for the condensate, non- condensate density, and
non-condensate correlation respectively. Fig. 7 shows the
corresponding plot at the finite temperature $T = 10 \hbar
\omega_{\rm trap} /k$. The main difference is that there are more
peaks appearing around frequency $0< \Omega_i < 1$, $i = 1,2$ than
at zero temperature.

Since the second order susceptibility $K^{(2)}(\Omega_1,\Omega_2)$
is by itself a product of several Green's functions ${\cal
U}(\omega)$ and linear susceptibilities $K^{(1)}(\omega)$, $\omega
= \Omega_1, \Omega_2$, and $\Omega_1+\Omega_2$, several features
of Figs. 6 and 7 are  closely related to the linear
susceptibility. From Fig. 1, it is clear that the linear
susceptibility near frequency of 1 dominates the spectrum at both
zero and finite temperatures. It is therefore expected that all
second order response $K^{(2)}(\Omega_1,\Omega_2)$ containing
$\Omega_1=1$ or $\Omega_2=1$ component in it will lend a much
stronger contribution to the spectrum than those not at
$\Omega_1=1$ or $\Omega_2=1$. This explains the features around
$\Omega_1=1$ and $\Omega_2=1$. In addition, the linear
susceptibility $K^{(1)}(\omega)$ at finite temperature has more
peaks around frequencies $0< \omega < 1$ than at zero temperature,
as there are more resonances below the frequency of 1 at finite
temperature. This leads to   the second order susceptibility
$K^{(2)}(\Omega_1,\Omega_2)$  displaying more peaks around
frequency $0< \Omega_1 < 1$ and $0< \Omega_2 < 1$, as  illustrated
in Fig. 7.

\section{Nonlinear Response of Fractional Quantum Hall Systems}

The Fractional Quantum Hall Effect (FQHE)~\cite{QHE} for a
two-dimensional ($2D$) electron gas in a strong, perpendicular,
external magnetic field is observed through the quantization of
the Hall dc conductivity $\sigma_{H}(\nu)$ as a function of the
filling factor $\nu \equiv (2\pi n)/( m \omega_{c})$, where $n$ is
the mean $2D$ density of the electrons, $\omega_{c} = eB/mc$ is
the cyclotron frequency, $B$ is the magnetic field; $e$ is the
electron charge (we set $c = 1$ and $\hbar = 1$). At $\nu =
1/(2k+1)$, where $k$ is an integer $k = 1, 2, \ldots$
$\sigma_{H}(\nu)$ varies in discrete steps, and is given by
$\sigma_{H}(\nu) = \nu(e^{2}/2\pi)$.

It has been shown that for $\nu = 1/(2k+1)$ the original $2D$
fermion problem can be mapped into a boson system coupled to a
Chern-Simons gauge field added to the time-independent external
magnetic field, where the original fermion system and the boson
system have the same charge density and Hall
conductivity~\cite{Zhang,Lopez,Meng}. The static Hall conductivity
($x y$ component of the conductivity tensor) calculated for this
boson system, $\sigma_{H} = \nu(e^{2}/2\pi)$, coincides with the
conductivity obtained by using Laughlin's ansatz for the
many-electron wavefunction~\cite{Laughlin}.  For an even inverse
filling number $\nu = 1/(2k)$ the original fermion problem is
mapped onto a composite fermion system coupled to a Chern-Simons
gauge field~\cite{Jain}. At $\nu = 1/2$ this gauge field
eliminates the external effective vector potential  for the
effective composite fermion system~\cite{HLR,Birman,Birman1}. In
this section we consider FQHE with the odd inverse denominator of
the filling number $\nu = 1/(2k+1)$.
 Using this mapping we can apply our
results based on a generalized coherent state (GCS) ansatz for the
many particle wave function of a weakly interacting effective
boson system~\cite{perelomov,Chernyak,choi} to compute the
second-order response of FQHE. Below we establish the
correspondence between the FQHE effective Bose Hamiltonian and the
Hamiltonian Eqs.~(\ref{hamb})-~(\ref{Vijkl}).

 We start with the many-electron effective Hamiltonian
$\hat {H}$ where the electron-electron interaction contains
Coulomb repulsion $V(\mathbf{r}-\mathbf{r}') =
1/|\mathbf{r}-\mathbf{r}'|$, external magnetic vector-potential
$\mathbf{A}(\mathbf{r})$, and an  external electrical scalar
potential $ V_{f}(\mathbf{r},t)$~\cite{Chernyak,choi}
\begin{eqnarray}
\label{hame} \hat{H} &=& \int d t \,\int d{\bf r}\,
\hat{\psi}_{e}^{\dagger}({\bf r},t) \left(
\frac{(-i\nabla_{\mathbf{r}} - e(\mathbf{A}(\mathbf{r})))^{2}}{2m}
- \mu\right) \hat{\psi}_{e}({\bf r},t) \nonumber \\    &+&
\frac{1}{2} \int d t \int d{\bf r}\,\int d{\bf
r}'\,\hat{\psi}_{e}^{\dagger}({\bf r},t)
\hat{\psi}_{e}^{\dagger}({\bf r}',t)V(\mathbf{r}-\mathbf{r}')
\hat{\psi}_{e}({\bf r}',t) \hat{\psi}_{e} ({\bf r},t)
 \nonumber \\    &+& \int d t \int d{\bf r}\,\hat{\psi}_{e}^{\dagger}({\bf
r},t) V_{f}(\mathbf{r},t) \hat{\psi}_{e} ({\bf r},t) ,
\end{eqnarray}
where $\hat{\psi}_{e}^{\dagger}({\bf r},t) $ and
$\hat{\psi}_{e}({\bf r},t) $ are Fermi creation and annihilation
operators; $m$ is the effective electron band mass and $\mu$ is
the chemical potential.

We next introduce a {\it quasiparticle} creation operator
$\hat{\psi}^{\dagger}({\bf r},t) $ which is related to
$\hat{\psi}_{e}^{\dagger}({\bf r},t)$ as\cite{HLR}
\begin{eqnarray}
\label{quasi} \hat{\psi}^{\dagger}({\bf r},t) &\equiv&
\hat{\psi}_{e}^{\dagger}({\bf r},t) \exp\left[-i \nu^{-1} \int d
{\bf r}' \arg({\bf r} - {\bf r}') \hat{n}({\bf r}',t) \right],
\end{eqnarray}
where $\arg({\bf r} - {\bf r}')$ is the angle between the vector
$({\bf r} - {\bf r}')$ and the direction of the Hall current
(which is perpendicular to both the external magnetic field
$\mathbf{B} = \mathbf{\nabla}_{\mathbf{r}} \times
\mathbf{A}(\mathbf{r})$ and the external electric field
$\mathbf{E} = - \nabla_{\mathbf{r}} V_{f}(\mathbf{r},t)$, which
are perpendicular to each other); $\nu^{-1}$ is an odd integer;
and $\hat{n}({\bf r},t)$ is the charge density operator, which is
the same for the actual Fermi and artificial Bose system
\begin{eqnarray}
\label{chden}\hat{n}({\bf r},t) \equiv
\hat{\psi}_{e}^{\dagger}({\bf r},t) \hat{\psi}_{e}({\bf r},t) =
\hat{\psi}^{\dagger}({\bf r},t)\hat{\psi}({\bf r},t).
\end{eqnarray}
The operators $\hat{\psi}^{\dagger}({\bf r},t)$ and
$\hat{\psi}({\bf r},t)$ satisfy Bose commutation
relations~\cite{Zhang,Lopez}
\begin{eqnarray}
\label{bosecom} && \hat{\psi}({\bf r})\hat{\psi}^{\dagger}({\bf
r}') - \hat{\psi}^{\dagger}({\bf r}') \hat{\psi}({\bf r}) = \delta
(\mathbf{r} - \mathbf{r}'); \nonumber \\
&& \hat{\psi}({\bf r})\hat{\psi}({\bf r}') -
\hat{\psi}({\bf r}') \hat{\psi}({\bf r}) = 0 ; \nonumber \\
&&  \hat{\psi}^{\dagger}({\bf r})\hat{\psi}^{\dagger}({\bf r}') -
\hat{\psi}^{\dagger}({\bf r}') \hat{\psi}^{\dagger}({\bf r}) = 0 .
\end{eqnarray}
Using the Bose quasiparticle operators Eq.~(\ref{quasi}), the
Hamiltonian Eq.~(\ref{hame}) can be written in a form of the
many-boson effective Hamiltonian $\hat {H}$ by adding the gauge
Chern-Simons vector potential
$\mathbf{a}(\mathbf{r})$~\cite{Zhang, Meng}
\begin{eqnarray}
\label{ham} \hat{H} &=& \int d t \,\int d{\bf r}\,
\hat{\psi}^{\dagger}({\bf r},t) \left(
\frac{(-i\nabla_{\mathbf{r}} - e(\mathbf{A}(\mathbf{r}) +
\mathbf{a}(\mathbf{r})))^{2}}{2m} - \mu\right) \hat{\psi}({\bf
r},t) \nonumber \\    &+& \frac{1}{2} \int d t \int d{\bf r}\,\int
d{\bf r}'\,\hat{\psi}^{\dagger}({\bf r},t)
\hat{\psi}^{\dagger}({\bf r}',t)V(\mathbf{r}-\mathbf{r}')
\hat{\psi}({\bf r}',t) \hat{\psi} ({\bf r},t)
 \nonumber \\    &+& \int d t \int d{\bf r}\,\hat{\psi}^{\dagger}({\bf
r},t) V_{f}(\mathbf{r},t) \hat{\psi} ({\bf r},t) .
\end{eqnarray}
In order for the static Hall conductivity of the boson Hamiltonian
Eq.~(\ref{ham}) in Hartree-Fock-Bogoliubov
approximation~\cite{ChernyakChoi} to coincide with the
conductivity obtained by using Laughlin's ansatz for the
many-electron wavefunction~\cite{Laughlin}, the magnetic and gauge
potentials in $k$ space should satisfy~\cite{Meng}
\begin{eqnarray}\label{gauge}
A_{\alpha}(k) + a_{\alpha}(k) = \frac{2\pi}{e\nu}\epsilon^{\alpha
\beta}\frac{ik_{\beta}}{k^{2}}(\hat{n}_{k} - n),
\end{eqnarray}
where $\epsilon^{\alpha \beta}$ is a unit antisymmetric tensor
\begin{equation} \label{eps}
\epsilon^{\alpha \beta} \equiv \left (
\begin{array}{cc}
0 & 1 \\
-1 & 0
\end{array}
\right );
\end{equation}
$n$ is the mean $2D$ electron density $n = (m
\nu\omega_{c})/(2\pi)$,  and the Fourier components of  the charge
density operator $\hat{n}_{k}$ with momentum $k$ is
\begin{eqnarray}\label{cdensity}
\hat{n}_{k} = \sum_{k} \hat{a}_{k}^{\dagger} \hat{a}_{k},
\end{eqnarray}
where $\hat{a}_{k}^{\dagger}$ and $\hat{a}_{k}$ are Fourier
components of  $\hat{\psi}^{\dagger}(\mathbf{r})$ and
$\hat{\psi}(\mathbf{r})$. Using a basis set of single electron
functions $\phi_{i}(\mathbf{r})$, the field operators may be
expanded in the form
\begin{eqnarray}\label{expbs1}
\hat \psi^\dagger(\mathbf{r},t) = \sum_{i}
\phi_{i}^{*}(\mathbf{r})\hat{a}_{i}^{\dagger}(t); \nonumber
\end{eqnarray}
\begin{eqnarray}\label{expbs2}
\hat \psi(\mathbf{r},t) = \sum_{i}
\phi_{i}(\mathbf{r})\hat{a}_{i}(t) ,
\end{eqnarray}
where $a_{i}^{\dagger}$ and $a_{i}$ are Bose operators.

We make the following GCS Hartree-Fock-Bogoliubov ansatz for the
time-dependent many-boson
wavefunction~~\cite{Blaizot_Ripka,ChernyakChoi}
\begin{eqnarray}
|\psi (t)\rangle &=& \exp \left( \int dt \int d\mathbf{r} \alpha
(\mathbf{r}, t) \hat{\psi}^{\dagger}(\mathbf{r}) +  \int dt \int
d\mathbf{r} d\mathbf{r}' \beta (\mathbf{r}, \mathbf{r}', t)
\hat{\psi}^{\dagger}(\mathbf{r})\hat{\psi}^{\dagger}(\mathbf{r}')
\right)
|\Omega _{0}\rangle \nonumber \\
&=& \exp \left( \sum_{i}\alpha _{i}(t)\hat{a}_{i}^{\dagger} +
\sum_{ij}\beta _{ij}(t)\hat{a}_{i}^{\dagger}\hat{a}_{j}^{\dagger}
\right) |\Omega _{0}\rangle , \label{ansatz}
\end{eqnarray}
where $| \Omega _{0}\rangle$ is an arbitrary normalized reference
state with $\langle \Omega _{0}|\Omega _{0}\rangle
=1$~\cite{ChernyakChoi}.

On the other hand, the operator set of GCS generators of the
closed algebra~\cite{perelomov,Chernyak} in the exponent of the
ansatz Eq.~(\ref{ansatz}) creates an extended Heisenberg-Weyl
algebra which may be obtained by a repeated application of the
standard boson commutators.

The Hamiltonian that describes the system of many-body interacting
bosons is obtained and comes from the basis set expansion of the
Hamiltonian Eq.~(\ref{ham}) and is given by~\cite{Meng}
\begin{equation}
\hat{H}= \sum_{ij}\hat{H}_{ij}\hat{a}_{i}^{\dagger }\hat{a}_{j} +
\sum_{ijkl}V_{ijkl}\hat{a}_{i}^{\dagger }\hat{a}_{j}^{\dagger
}\hat{a}_{k}\hat{a}_{l} + \eta \sum_{ij} E_{ij}(t)
\hat{a}^{\dagger}_{i}\hat{a}_{j},  \label{hamiltonian}
\end{equation}
Note that the Hamiltonian Eq.~(\ref{hamiltonian}) coincides with
Eqs.~(\ref{hamb})-~(\ref{Vijkl}) provided we replace in
Eqs.~(\ref{Hamiltonian}) $H_{ij}^{sp}$ by $H_{ij}$. In
Eq.~(\ref{hamiltonian}) the single-electron matrix element,
$H_{ij}$ given by
\begin{equation}\label{eq-1electr}
H_{ij} = \int d{\bf r}\, \phi_i^* ({\bf r})
   \left( \frac{(-i\nabla_{\mathbf{r}} - e(\mathbf{A}(\mathbf{r}) + \mathbf{a}(\mathbf{r})))^{2} }{2m} - \mu\right)
   \phi_j ({\bf r}),
\end{equation}
$V_{ijkl}$ is a Coulomb repulsion between two electrons (we put
electron charge $e = 1$)
\begin{equation}\label{eq-2electrV}
V_{ijkl} = \int d{\bf r}_1 d{\bf r}_2\, \phi_i^* ({\bf r}_1)
\phi_j^* ({\bf r}_2) \frac{1}{| {\bf r}_1 - {\bf r}_2 |}
   \phi_k ({\bf r}_1) \phi_l ({\bf r}_2) ,
\end{equation}
and the external electrical field $E(t)$ field expanded in a basis
set is
\begin{equation}\label{eq-2electrVf}
\sum_{ij} E_{ij}(t) \hat{a}^{\dagger}_{i}\hat{a}_{j} = \int \!
d{\bf r} \, \phi^{*}_{i}({\bf r}) V_{f}({\bf r},t) \phi_{j}({\bf
r})  \hat{a}^{\dagger}_{i}\hat{a}_{j} .
\end{equation}
We can, therefore, apply the results obtained for the Hamiltonian
Eqs.~(\ref{hamb})-~(\ref{Vijkl}) to the  FQHE by simply replacing
$H_{ij}^{sp}$ in Eqs.~(\ref{Hamiltonian})  by $H_{ij}$
Eq.~(\ref{eq-1electr}).

Since in a homogeneous system the momentum is conserved, we can
use the plane wave basis, i.e. the eigenfunctions of a momentum
$\mathbf{p}$ ($\phi_{\mathbf{p}}(\mathbf{r}) =
S^{-1/2}\exp(-i\mathbf{p}\mathbf{r})$; where $S$ is the $2D$
volume of a system) by replacing the indices $i$ by the momentum
$\mathbf{p}$.

The second-order response is given by Eqs.~(D1)-~(D3),(D24) in the
plane wave basis, provided we use  for  the Hamiltonian
Eqs.(1)-~(2) in the Liouvillian Eq.~(B6) with $V_{trap} = 0$, and
replace the inter-particle interaction $V_{ijkl}$ by
\begin{eqnarray}
\label{Vijklimp} V_{\mathbf{p} - \mathbf{p}'} = \frac{2\pi e^{2}}{
\epsilon |\mathbf{p} - \mathbf{p}'| } +
\frac{4\pi^{2}n}{m\nu^{2}|\mathbf{p} - \mathbf{p}'|^{2}},
\end{eqnarray}
where the first term in the r.h.s. represents the $2D$ Fourier
transform component of the Coulomb repulsion and the second term
comes from substituting Eq.~(\ref{gauge}) in
Eq.~(\ref{hamiltonian}). With these substitutions, the FQHE boson
Hamiltonian Eq.~(\ref{hamiltonian}) reduced to the classical form
Eq.~(19)~\cite{Meng} and is equivalent to the classical boson
Hamiltonian (Eq.~(40) in~\cite{ChernyakChoi}). The double
excitations are given by the eigenvalues of the Liouvillian
Eq.~(B6). The anisotropy in the gauge term Eq.~(\ref{gauge}) in
the Hamiltonian Eq.~(\ref{hamiltonian}) gives a non-vanishing
second-order response, shifting the double-excitation energies
with respect to twice the single excitations, as is the case in
isotropic systems.

\section{Conclusions}

We have calculated the second order mechanical response functions
and susceptibilities of both zero and finite temperature BEC. The
systematic, perturbative solution of the TDHFB equations for
trapped, atomic BEC enables us to analyze the dynamics of finite
temperature BEC. The response of both the non-condensate atoms as
well as the condensate were calculated. The calculations apply for
a general  perturbation of the form $\sum_{ij} \int \! d^{3}{\bf
r} \, \phi^{*}_{i}({\bf r}) V_{f}({\bf r},t) \hat{a}_{i}^{\dagger}
\hat{a}_{j}$, where the shape of the external force $V_{f}({\bf
r},t)$ is left arbitrary. The calculated second order response
functions were found to show a strong dependence on position and
temperature. In the frequency domain, we have observed distinct
nonlinear mixing effect, which displays enhanced response at a
number of $\Omega_1$-$\Omega_2$ combinations. The application of
the GCS ansatz for the derivation of the second-order response of
FQHE $2D$ electron liquid was presented.

The second order response for BEC has not been reported yet, and
most related work is limited to the linear response. As in
nonlinear optics, the second and higher order response functions
and susceptibilities are clearly expected to be indispensable in
characterizing the BEC dynamics in the presence of strong
perturbations, and more importantly, in further development of
matter-wave nonlinear optics~\cite{Band}.

\acknowledgments
The support of NSF Grant No. CHE-0132571 is gratefully acknowledged.

\appendix

\section{The TDHFB Equations} \label{TDHFBEqs}

The TDHFB equations of motion that couples $z$, $\rho$ and
$\kappa$ are~\cite{choi,Prou2}
\begin{eqnarray}
i \hbar \frac{d z}{dt} & = & \left [ {\cal H}_{z} + \eta E(t) \right ] z +
{\cal  H}_{z*} z^{*} ,   \label{zdot} \\
i \hbar \frac{d \rho}{dt} & = & [h, \rho] - (\kappa \Delta^{*} - \Delta
\kappa^{*}) + \eta [E(t), \rho ] ,  \label{rhodot} \\
i \hbar \frac{d \kappa}{dt} & = & (h \kappa + \kappa h^{*}) + (\rho \Delta +
\Delta \rho^{*}) + \Delta + \eta [E(t),\kappa]_{+} ,  \label{kappadot}
\end{eqnarray}
where $[ \dots ]_{+}$ denotes the anticommutator.
Here,  ${\cal H}_{z}$ , ${\cal H}_{z*}$, $h$, and $\Delta$ are $N \times N$
matrices  with $n$ being the number of basis wave functions used:
\begin{eqnarray}
\left [ {\cal H}_{z} \right ]_{i,j} & = & H_{ij} - \mu + \sum_{kl} V_{iklj}
\left [
z^{*}_{k}z_{l} + 2 \rho_{lk} \right ] \label{h_z} \\
\left [ {\cal H}_{z*} \right ]_{i,j} & = & \sum_{kl} V_{ijkl} \kappa_{kl}
\label{h_z_star} \\
h_{ij} & = & H_{ij} - \mu + 2 \sum_{kl} V_{iklj}  \left [
z_{k}^{*}z_{l} + \rho_{lk} \right ], \label{h} \\
\Delta_{ij} & = & \sum_{kl}  V_{ijkl}  \left[ z_{k} z_{l}
+ \kappa_{kl} \right]~.  \label{Delta}
\end{eqnarray}
$h$ is known as the ``Hartree-Fock Hamiltonian''and $\Delta$ as
the ``pairing field''~\cite{Blaizot_Ripka}. $\mu$ is the chemical
potential introduced in the Hamiltonian, Eq. (\ref{Hamiltonian}).

\section{Matrix \mbox{${\cal L}$} and vector $\zeta$} \label{Lmatrix}

In this Appendix we define the $2N(2N+1) \times 2N(2N+1)$  Liouvillian
matrix of Eq. (\ref{general}).
As discussed in the main text, it suffices to define the zero'th and the
first order matrices
${\cal L}^{(0)}$ and ${\cal L}^{(1)}$ only, since ${\cal L}^{(n)}$ for all $n
\geq 1$ is identical.
First of all, ${\cal L}^{(0)}$ is defined as follows:
\begin{equation}
{\cal L}^{(0)} = \left (
\begin{array}{cccccc}
{\cal H}_{z}^{(0)} - \mu & {\cal H}_{z*}^{(0)} & 0 & 0 & 0 & 0 \\
{\cal H}_{z*}^{(0)*} & {\cal H}_{z}^{(0)*} - \mu & 0 & 0 & 0 & 0 \\
0 & 0 & {\cal H}^{(-)} & {\cal D}^{\Delta} & 0 & {\cal D} \\
0 & 0 & -{\cal D}^{\Delta *} & {\cal H}^{(+)} & {\cal D} & 0 \\
0 & 0 & 0 & {\cal D}^{*} & {\cal H}^{(-)*} & {\cal D}^{\Delta *} \\
0 & 0 & {\cal D}^{*} & 0 & -{\cal D}^{\Delta} & {\cal H}^{(+)*}
\end{array}
\right ), \label{SCTIHFB}
\end{equation}
where ${\cal H}_{z}$ and ${\cal H}_{z*}$ of ${\cal L}^{(0)}$ are as defined in
Eqs. (\ref{h_z}-\ref{h_z_star}) and  the remaining $N^{2} \times N^{2}$
submatrices are defined as:
\begin{eqnarray}
{\cal H}^{(-)}_{ij, mn} & = & h^{(0)}_{im} \delta_{jn} - h^{(0)}_{nj}
\delta_{im} \\
{\cal H}^{(+)}_{ij, mn} & = & h^{(0)}_{im} \delta_{jn} + h^{(0)}_{nj}
\delta_{im} + V_{ijmn} \\
{\cal D}_{ij, mn} & = & \Delta^{(0)}_{im} \delta_{jn} \\
{\cal D}_{ij, mn}^{\Delta} & = & -\Delta_{nj}^{*(0)} \delta_{im}
\end{eqnarray}
with $h^{(0)}$ and $\Delta^{(0)}$ being the Hartree-Fock
Hamiltonian and the pairing field defined in Eqs. (\ref{h}-\ref{Delta}).

For higher orders  $n \geq 1 $ (in particular, $n = 2$),
\begin{equation}
{\cal L}^{(n)} \equiv {\cal L}^{(0)} + {\cal L}',
\end{equation}
where ${\cal L}^{(0)}$ is the zero'th order matrix defined above, and
\begin{equation}
{\cal L}' = \left (
\begin{array}{cccccc}
{\cal V}^{zz1} & {\cal V}^{zz2} & {\cal V}^{z1} & {\cal V}^{z2} & 0 & 0 \\
{\cal V}^{zz2*} & {\cal V}^{zz1*} & 0 & 0 & {\cal V}^{z1*} & {\cal V}^{z2*}
\\
{\cal V}^{\rho z1} & {\cal V}^{\rho z2} & {\cal W}^{\rho h} & {\cal W}%
^{\kappa \Delta} & 0 & {\cal W}^{\kappa \Delta \dagger} \\
{\cal V}^{\kappa z1} & {\cal V}^{\kappa z2} & {\cal W}^{\kappa h} & {\cal
W}%
^{\rho \Delta} & {\cal W}^{\kappa h \dagger} & 0 \\
{\cal V}^{\rho z2*} & {\cal V}^{\rho z1*} & 0 & ({\cal W}^{\kappa \Delta
\dagger})^{*} & ({\cal W}^{\rho h})^{*} & ({\cal W}^{\kappa \Delta})^{*} \\
{\cal V}^{\kappa z2*} & {\cal V}^{\kappa z1*} & ({\cal W}^{\kappa h
\dagger})^{*} & 0 & ({\cal W}^{\kappa h})^{*} & ({\cal W}^{\rho \Delta})^{*}
\end{array}
\right ).
\end{equation}
The set of $N \times N$ submatrices ${\cal V}^{zz1}$ and ${\cal V}^{zz2}$,
$N
\times N^{2}$ submatrices  ${\cal V}^{z1}$ and ${\cal V}^{z2}$,
$N^2 \times N$ submatrices
${\cal V}^{\rho z1}$,  ${\cal V}^{\rho z2}$,  ${\cal V}^{\kappa z1}$, and
${\cal V}^{\kappa z2}$, and $N^{2} \times N^{2}$ component submatrices
${\cal W}^{\rho h}$, ${\cal W}^{\rho \Delta}$, ${\cal W}^{\kappa h}$, and
${\cal W}^{\kappa \Delta}$ of ${\cal L}'$ are given as follows:
\begin{eqnarray}
{\cal V}^{zz1}_{i,l} & = & \sum_{kr} V_{iklr} z^{*(0)}_{k}z^{(0)}_{r}
\;\;\;\;\;
{\cal V}^{zz2}_{i,k}  =  \sum_{lr} V_{iklr} z^{(0)}_{l}z^{(0)}_{r} \\
{\cal V}^{z1}_{i, kl} & = & 2 \sum_{r} V_{ilkr} z^{(0)}_{r}  \;\;\;\;\;
{\cal
V}^{z2}_{i, kl}  =  \sum_{r} V_{iklr} z^{(0)}_{r} \\
{\cal V}^{\rho z1}_{ij,l} & = & 2 \sum_{kr} V_{iklr} z^{*(0)}_{k}
\rho^{(0)}_{rj} - V_{rklj} z^{*(0)}_{k} \rho^{(0)}_{ir} + \sum_{kr} \left [
V_{irkl} z^{(0)}_{k} + V_{irlk} z^{(0)}_{k} \right ] \kappa^{*(0)}_{rj} \\
{\cal V}^{\rho z2}_{ij,k} & = & 2 \sum_{lr} V_{iklr} z^{(0)}_{l}
\rho^{(0)}_{rj} - V_{rklj} z^{(0)}_{l} \rho^{(0)}_{ir} - \sum_{lr} \left [
V_{rjkl} z^{*(0)}_{l} + V_{rjlk} z^{*(0)}_{l} \right ] \kappa^{(0)}_{ir} \\
{\cal V}^{\kappa z1}_{ij,k} & = & 2 \sum_{lr} V_{ilkr} z^{*(0)}_{l}
\kappa^{(0)}_{rj} + V_{rklj} z^{*(0)}_{l} \kappa^{(0)}_{ir} + \sum_{lr}
\left [ V_{rjkl} z^{(0)}_{l} + V_{rjlk} z^{(0)}_{l} \right ] \rho^{(0)}_{ir}
\nonumber \\
& + & \sum_{lr} \left [ V_{irkl} z^{(0)}_{l} + V_{irlk} z^{(0)}_{l} \right ]
\rho^{*(0)}_{rj} + \sum_{l} \left [ V_{ijkl} z^{(0)}_{l} + V_{ijlk}
z^{(0)}_{l} \right ] \\
{\cal V}^{\kappa z2}_{ij,k} & = & 2 \sum_{lr} V_{iklr} z^{(0)}_{l}
\kappa^{(0)}_{rj} + V_{rlkj} z^{(0)}_{l} \kappa^{(0)}_{ir} \\
{\cal W}^{\rho h}_{ij, kl} & = & 2 \sum_{r} V_{iklr} \rho^{(0)}_{rj} -
V_{rklj} \rho^{(0)}_{ir}  \;\;\;\;\;
{\cal W}^{\rho \Delta}_{ij, kl}  =  \sum_{r} V_{irkl} \rho^{(0)*}_{rj} +
V_{rjkl} \rho^{(0)}_{ir} \\
{\cal W}^{\kappa h}_{ij, kl} & = & \sum_{r} V_{iklr} \kappa^{(0)}_{rj}
\;\;\;\;\; {\cal W}^{\kappa h \dagger}_{ij, kl}  =  \sum_{r} V_{rklj}
\kappa^{(0)}_{ir} \\
{\cal W}^{\kappa \Delta}_{ij, kl} & = & \sum_{r} V_{irkl} \kappa^{(0)*}_{rj}
\;\;\;\;\;
{\cal W}^{\kappa \Delta \dagger}_{ij, kl}  =  \sum_{r} V_{rjkl}
\kappa^{(0)}_{ir}
\end{eqnarray}

In addition, the $2N(2N+1) \times 1$ vector $\zeta(t)$ is defined as
follows:
\begin{equation}
\zeta(t) \equiv \left (
\begin{array}{cccccc}
E(t) & 0 & 0 & 0 & 0 & 0\\
0 & E^{*}(t) & 0 & 0 & 0 & 0\\
0 & 0 & \epsilon^{(-)}(t) & 0 & 0 & 0\\
0 & 0 & 0 & \epsilon^{(+)}(t) & 0 & 0 \\
0 & 0 & 0 &  0 & \left [\epsilon^{(-)}(t) \right ]^{*} & 0 \\
0 & 0 & 0 &  0 & 0 & \left [ \epsilon^{(+)}(t) \right ]^{*}
\end{array}
\right ) \left (
\begin{array}{c}
\vec{z}^{(0)} \\
\vec{z}^{(0)*} \\
\vec{\rho}^{(0)} \\
\vec{\kappa}^{(0)} \\
\vec{\rho}^{(0)*} \\
\vec{\kappa}^{(0)*}
\end{array}
\right ),
\end{equation}
where $E(t)$ is as defined in Eq. (\ref{E_t}), and we have further defined the $N^2 \times N^2$ submatrices
\begin{equation}
{\cal \epsilon}^{(\pm)}(t)_{ij, kl}  =  E_{ik}(t) \delta_{jl} \pm
E_{lj}(t) \delta_{ik} .
\end{equation}

\section{Second order response functions} \label{K2}

In this Appendix, we first summaries the final result for the
response functions, and then provide a more detailed derivation.
In the last subsection of this Appendix, we also show an
alternative form for the response functions written in the basis
of the eigenvectors of the Liouvillian.

\subsection{Final expression}
The second order response  functions for the condensate, the non-condensate
density and correlation that we use in our numerical calculations are given by Eqs. (\ref{K2tz}) - (\ref{K2tkappa}) as follows:
\begin{equation}
K^{(2)}_{z}(t,t_1, t_2, {\bf r},{\bf r}_1,{\bf r}_2) = \sum_{i =1}^{N}
\left [\tilde{\Upsilon}({\bf r}) \left ( \vec{K}^{(2)}_{I}(t,t_1,t_2,{\bf
r}_1,
{\bf r}_2) + \vec{K}^{(2)}_{II}(t,t_1,t_2,{\bf r}_1,{\bf r}_2) \right )
\right
]_{i} \label{K2tz}
\end{equation}
\begin{equation}
K^{(2)}_{\rho}(t, t_1, t_2, {\bf r},{\bf r}_1,{\bf r}_2) = \sum_{i = 2N +
1}^{2N + N^2} \left [\tilde{\Upsilon}({\bf r}) \left (
\vec{K}^{(2)}_{I}(t,t_1,t_2,
{\bf r}_1,{\bf r}_2) + \vec{K}^{(2)}_{II}(t,t_1,t_2,{\bf r}_1,{\bf r}_2)
\right
)
\right ]_{i} \label{K2trho}
\end{equation}
\begin{equation}
K^{(2)}_{\kappa}(t,t_1, t_2, {\bf r},{\bf r}_1,{\bf r}_2) = \sum_{i = 2N+N^2
+ 1}^{2N+ 2N^2} \left [\tilde{\Upsilon}({\bf r}) \left (
\vec{K}^{(2)}_{I}(t,t_1,t_2,{\bf r}_1,{\bf r}_2) +
\vec{K}^{(2)}_{II}(t,t_1,t_2,{\bf r}
_1,{\bf r}_2) \right ) \right ]_{i}  \label{K2tkappa}
\end{equation}
where $2N(2N+1) \times 2N(2N+1)$ matrix $\tilde{\Upsilon}({\bf
r})$  was defined in Eq. (\ref{upsilontilde}), and  the $2N(2N+1)
\times 1$ vectors $\vec{K}^{(2)}_{I}$ and $\vec{K}^{(2)}_{II}$ are
defined as follows:
\begin{equation}
\vec{K}^{(2)}_{I}(t,t_1,t_2,{\bf r}_1,{\bf r}_2)  =  {\cal U}(t-t_1)
\tilde{\Phi}({\bf r}_1){\cal U}(t_1-t_2) \tilde{\Phi}({\bf
r}_2)\vec{\psi}^{(0)}, \label{KIt1}
\end{equation}
\begin{equation}
\vec{K}^{(2)}_{II}(t,t_1,t_2,{\bf r}_1,{\bf r}_2) =  \int_{0}^{t} d \tau
{\cal
U}(t-\tau) \vec{\Xi}_{K}(\tau-t_1,\tau - t_2, {\bf r}_1,{\bf r}_2),
\label{KIIt1}
\end{equation}
where
\begin{equation}
{\cal U}(t-t^{\prime })   \equiv   \exp \left[ -\frac{i}{\hbar }{\cal L}^{(2)}(t-
t^{\prime }) \right], \label{U}
\end{equation}
and
\begin{equation}
\tilde{\Phi}({\bf r}^{\prime })   =  {\rm diag} \left[ \Phi ({\bf r}^{\prime
}),\Phi ^{\ast }({\bf r}^{\prime }),\Phi ^{(-)}( {\bf r}^{\prime }),\Phi
^{(+)}({\bf r}^{\prime }),\Phi ^{(-)\ast }({\bf r}^{\prime }),\Phi ^{(+)\ast
}({\bf r}^{\prime })\right]. \label{phitilde}
\end{equation}
$\tilde{\Phi}({\bf r}^{\prime })$ is therefore a $2N(2N+1) \times 2N(2N+1)$
block diagonal square matrix with the blocks consisting of $N \times N$
square matrices
\begin{equation}
\left[ \Phi ({\bf r})\right] _{ij}=\phi _{i}^{\ast }({\bf r})\phi _{j}({\bf
r})
\end{equation}
and $N^{2}\times N^{2}$ square matrices
\begin{equation}
\left[ \Phi ^{(\pm )}({\bf r})\right] _{ij,mn}=\phi _{i}^{\ast }({\bf
r})\phi
_{m}({\bf r})\delta _{jn}\pm \phi _{n}^{\ast }({\bf r})\phi _{j}({\bf
r})\delta
_{im}.
\end{equation}

The vector $\vec{\Xi}_{K}$ of Eq. (\ref{KIIt1}) may be written
\begin{equation}
\vec{\Xi}_{K} = [{\cal Z}_{K}, {\cal Z}^{*}_{K}, {\cal R}_{K}, {\cal K}_{K},
{\cal R}^{*}_{K}, {\cal K}^{*}_{K}]^{T} \label{XiK}
\end{equation}
with the $N \times 1$ matrix ${\cal Z}_{K}$ and $N^{2} \times 1$ matrices ${\cal R}_{K}$, ${\cal K}_{K}$ given as follows:
\begin{eqnarray}
\left [ {\cal Z}_{K} \right ]_{i} & = & \sum_{klr} V_{iklr} \left
[ [\vec{K}_z^{(1)*}(\tau - t_1, {\bf r}_{1})]_{k} z^{(0)}_{l}
[\vec{K}_z^{(1)} (\tau - t_2, {\bf r}_{2})]_{r} \right. \nonumber \\
& + & [\vec{K}_z^{(1)*}(\tau - t_1, {\bf
r}_{1})]_{k}[\vec{K}_z^{(1)}(\tau - t_2, {\bf r}_{2})]_{l}
z^{(0)}_{r}  +  z^{*(0)}_{k}[\vec{K}_z^{(1)}(\tau - t_1, {\bf
r}_{1})]_{l}
[\vec{K}_z^{(1)}(\tau - t_2, {\bf r}_{2})]_{r} \nonumber \\
&+& 2 [\vec{K}_\rho^{(1)}(\tau - t_1, {\bf r}_{1})]_{lk}
[\vec{K}_z^{(1)}(\tau - t_2, {\bf r}_{2})]_{r}  +  \left.
[\vec{K}_\kappa^{(1)}(\tau - t_1, {\bf
r}_{1})]_{kl}[\vec{K}_z^{(1)}(\tau - t_2, {\bf r}_{2})]_{r}
\right ] \label{XiKz} \\
\left [ {\cal R}_{K} \right ]_{ij}  & = & 2 \sum_{rkl} V_{iklr}
[\vec{K}_\rho^{(1)}(\tau - t_1, {\bf
r}_{1})]_{rj}[\vec{K}_\rho^{(1)}(\tau - t_2, {\bf r}_{2})]_{kl}
\nonumber \\
& - & V_{rklj} [\vec{K}_\rho^{(1)}(\tau - t_1, {\bf
r}_{1})]_{kl}[\vec{K}_\rho^{(1)}(\tau - t_2, {\bf r}_{2})]_{ir}
 +  \sum_{rkl} V_{irkl} [\vec{K}_\kappa^{(1)}(\tau - t_1, {\bf
r}_{1})]_{kl}[\vec{K}_\kappa^{*(1)}(\tau - t_2, {\bf r}_{2})]_{rj}
\nonumber \\
&+& V_{rjkl} [\vec{K}_\kappa^{*(1)}(\tau - t_1, {\bf
r}_{1})]_{kl}[\vec{K}_\kappa^{(1)}(\tau
- t_2, {\bf r}_{2})]_{ir}  \nonumber \\
& + & 2 \sum_{rkl} V_{iklr} [\vec{K}_z^{*(1)}(\tau - t_1, {\bf
r}_{1})]_{k}[\vec{K}_z^{(1)}(\tau - t_2, {\bf r}_{2})]_{l}
\rho^{(0)}_{rj} \nonumber \\
& - & V_{rklj} [\vec{K}_z^{*(1)}(\tau - t_1, {\bf r}_{1})]_{k}[\vec{K}_z^{(1)}(\tau - t_2, {\bf r}_{2})]_{l} \rho^{(0)}_{ir}  \nonumber \\
& + & 2 \sum_{rkl} V_{iklr} \left [ [\vec{K}_z^{*(1)}(\tau - t_1,
{\bf r}_{1})]_{k}z^{(0)}_{l} + z^{*(0)}_{k}[\vec{K}_z^{(1)}(\tau -
t_1, {\bf r}_{1})]_{l} \right ] [\vec{K}_\rho^{(1)}(\tau - t_2,
{\bf r}_{2})]_{rj} \nonumber \\
& - &  V_{rklj} \left [ [\vec{K}_z^{*(1)}(\tau - t_1, {\bf
r}_{1})]_{k}z^{(0)}_{l} + z^{*(0)}_{k}[\vec{K}_z^{(1)}(\tau - t_1,
{\bf r}_{1})]_{l} \right ] [\vec{K}_\rho^{(1)}(\tau - t_2, {\bf
r}_{2})]_{ir}
\nonumber \\
& + & \sum_{rkl} V_{rjkl} \left [ [\vec{K}_z^{(1)*}(\tau - t_1,
{\bf r}_{1})]_{k}z^{*(0)}_{l} + z^{*(0)}_{k}[\vec{K}_z^{*(1)}(\tau
- t_1, {\bf r}_{1})]_{l} \right ]
[\vec{K}_\kappa^{(1)}(\tau - t_2, {\bf r}_{2})]_{ir} \nonumber \\
& - & V_{irkl} \left [ [\vec{K}_z^{(1)}(\tau - t_1, {\bf
r}_{1})]_{k}z^{(0)}_{l} + z^{(0)}_{k}[\vec{K}_z^{(1)}(\tau - t_1,
{\bf r}_{1})]_{l} \right ] [\vec{K}_\kappa^{*(1)}(\tau - t_2, {\bf
r}_{2})]_{rj}
\nonumber \\
& + & \sum_{rkl} V_{rjkl} [\vec{K}_z^{(1)}(\tau - t_1, {\bf
r}_{1})]_{k}[\vec{K}_z^{(1)}(\tau - t_2, {\bf r}_{2})]_{l}
\kappa^{(0)}_{ir} \nonumber \\
& - & V_{irkl} [\vec{K}_z^{(1)}(\tau - t_1, {\bf
r}_{1})]_{k}[\vec{K}_z^{(1)}(\tau -
t_2, {\bf r}_{2})]_{l} \kappa^{*(0)}_{rj} \label{XiKrho} \\
\left [ {\cal K}_{K} \right ]_{ij} & = & 2 \sum_{rkl} V_{iklr}
[\vec{K}_\kappa^{(1)}(\tau - t_1, {\bf
r}_{1})]_{rj}[\vec{K}_\rho^{(1)}(\tau - t_2, {\bf r}_{2})]_{lk}
\nonumber \\
&+& V_{rklj} [\vec{K}_\rho^{(1)}(\tau - t_1, {\bf
r}_{1})]_{kl}[\vec{K}_\kappa^{(1)}(\tau - t_2, {\bf r}_{2})]_{ir}
 +  \sum_{rkl} V_{irkl} [\vec{K}_\kappa^{(1)}(\tau - t_1, {\bf
r}_{1})]_{kl}[\vec{K}_\rho^{(1)*}(\tau - t_2, {\bf r}_{2})]_{rj}
\nonumber \\
&+& V_{rjkl} [\vec{K}_\kappa^{(1)}(\tau - t_1, {\bf
r}_{1})]_{kl}[\vec{K}_\rho^{(1)}(\tau - t_2, {\bf r}_{2})]_{ir}
 +  2 \sum_{rkl} V_{iklr} [\vec{K}_z^{*(1)}(\tau - t_1, {\bf
r}_{1})]_{k}[\vec{K}_z^{(1)}(\tau - t_2, {\bf r}_{2})]_{l}
\kappa^{(0)}_{rj} \nonumber \\
& + & V_{rklj} [\vec{K}_z^{(1)}(\tau - t_1, {\bf
r}_{1})]_{k}[\vec{K}_z^{*(1)}(\tau -
t_2, {\bf r}_{2})]_{l} \kappa^{(0)}_{ir}  \nonumber \\
& + & 2 \sum_{rkl} V_{iklr} \left [ [\vec{K}_z^{(1)*}(\tau - t_1,
{\bf r}_{1})]_{k}z^{(0)}_{l} + z^{*(0)}_{k}[\vec{K}_z^{(1)}(\tau -
t_1, {\bf r}_{1})]_{l} \right
][\vec{K}_\kappa^{(2)}(\tau - t_1, {\bf r}_{2})]_{rj} \nonumber \\
& + & V_{rklj} \left [ [\vec{K}_z^{(1)}(\tau - t_1, {\bf
r}_{1})]_{k}z^{*(0)}_{l} + z^{(0)}_{k}[\vec{K}_z^{*(1)}(\tau -
t_1, {\bf r}_{1})]_{l} \right ] [\vec{K}_\kappa^{(1)}(\tau - t_2,
{\bf r}_{2})]_{ir}
\nonumber \\
& + & \sum_{rkl} V_{rjkl} \left [ [\vec{K}_z^{(1)}(\tau - t_1,
{\bf r}_{1})]_{k}z^{(0)}_{l} + z^{(0)}_{k}[\vec{K}_z^{(1)}(\tau -
t_1, {\bf r}_{1})]_{l} \right ]
[\vec{K}_\rho^{(1)}(\tau - t_2, {\bf r}_{2})]_{ir}  \nonumber \\
& + & V_{irkl} \left [ [\vec{K}_z^{(1)}(\tau - t_1, {\bf
r}_{1})]_{k}z^{(0)}_{l} + z^{(0)}_{k}[\vec{K}_z^{(1)}(\tau - t_1,
{\bf r}_{1})]_{l} \right ] [\vec{K}_\rho^{*(1)}(\tau - t_2, {\bf
r}_{2})]_{rj}
\nonumber \\
& + & \sum_{rkl} V_{rjkl} [\vec{K}_z^{(1)}(\tau - t_1, {\bf
r}_{1})]_{k}[\vec{K}_z^{(1)}(\tau - t_2, {\bf r}_{2})]_{l}
\rho^{(0)}_{ir} \nonumber \\
& + & V_{irkl} [\vec{K}_z^{(1)}(\tau - t_1, {\bf
r}_{1})]_{k}[\vec{K}_z^{(1)}(\tau - t_2, {\bf
r}_{2})]_{l} \rho^{*(0)}_{rj} \nonumber \\
& + & \sum_{kl} V_{ijkl} [\vec{K}_z^{(1)}(\tau - t_1, {\bf
r}_{1})]_{k}[\vec{K}_z^{(1)}(\tau - t_2, {\bf r}_{2})]_{l},
\label{XiKkappa}
\end{eqnarray}
where the vectors $\vec{K}^{(1)}_{z}(\tau - t_i, {\bf r}_i)$,
$\vec{K}^{(1)}_{\rho}(t-
t_1, {\bf r}_i)$ and $\vec{K}^{(1)}_{\kappa}(\tau - t_i, {\bf r}_i)$ found in Eqs. (\ref{XiKz}-\ref{XiKkappa}) are defined
as
\begin{equation}
\vec{K}^{(1)}_{\alpha}(\tau - t_i,{\bf r}_i) =  {\cal U}^{(\alpha)}(\tau - t_i)
\tilde{\Phi}({\bf r}_i) \vec{\psi}^{(0)}, \;\;\;\; \alpha = z, \rho, \kappa, \;\;\;\; i = 1,2. \label{K_zetc}
\end{equation}
${\cal U}^{(\alpha)}(\tau - t_i)$ are the
submatrices of ${\cal U}(t - t_1)$ such that
\begin{equation}
{\cal U}(t -t_1) = \left [ {\cal U}^{(z)}(\tau - t_i), {\cal U}^{(z)*}(\tau - t_i),
{\cal U}^{(\rho)}(\tau - t_i), {\cal U}^{(\kappa)}(\tau - t_i), {\cal
U}^{(\rho)*}(\tau - t_i),
{\cal U}^{(\kappa)*}(\tau - t_i) \right ]^{T} ,
\end{equation}
where ${\cal U}^{(z)}(\tau - t_i)$ is an $N \times 2N(2N + 1)$ submatrix while $%
{\cal U}^{(\gamma)}(\tau - t_i)$, $\gamma = \rho, \kappa, \rho^{*}, \kappa^{*}$
is an
$N^2 \times 2N(2N+1)$ submatrix, i.e. the submatrix ${\cal U}^{(z)}(\tau - t_i)$
is
stacked on top of submatrix ${\cal U}^{(z)*}(\tau - t_i)$ which, in turn, is
stacked
on top of submatrices ${\cal U}^{(\gamma)}(\tau - t_i)$. It is to be noted that
$\vec{K}^{(1)}_{\alpha}(\tau - t_i, {\bf r}_i)$ as defined here are $N \times 1$
and
$N^2 \times 1$ {\em vectors}, not scalar quantities obtained by integrating
the
scalar function $K^{(1)}_{\alpha}(\tau - t_i, {\bf r}, {\bf r}_i)$ over ${\bf
r}$.

\subsection{Derivation}

Writing the second order solution to TDHFB explicitly, we have
\begin{eqnarray}
\vec{\psi}^{(2)}(t) & =  & \frac{1}{i \hbar} \int_{0}^{t} \exp
\left [ -
\frac{i}{\hbar}{\cal L}^{(2)}(t-t_1) \right ] \vec{\Gamma}(t_{1}) dt_1 \\
& \equiv &    \frac{1}{i \hbar} \int_{0}^{t} {\cal U}(t - t_1)  \vec{\Gamma}(t_{1}) dt_1   \label{secondsol}
\end{eqnarray}
where
\begin{equation}
\vec{\Gamma}(t_{1})  =  \zeta(t_1) \vec{\psi}^{(1)}(t_1) + \vec{\Xi}(t_1) .
\label{gamma}
\end{equation}
i.e. for the second order response, $\lambda^{(2)}(t) \equiv \zeta(t)
\vec{\psi}^{(1)}(t) + \vec{\Xi}(t)$  in Eq. (\ref{general}) where $\zeta(t)$
is
given in Appendix \ref{Lmatrix}, and  $\vec{\Xi}(t)$ is a $2N(2N+1)
\times 1$ vector originating from the terms in the expansion which are made
up
of
products of two first order variables i.e. $z^{(1)}$, $\rho^{(1)}$ and
$\kappa^{(1)}$. The vector $\vec{\Xi}(t)$ can be written as $\vec{\Xi}(t) =
[{\cal Z%
}, {\cal Z}^{*} {\cal R}, {\cal K}, {\cal R}^{*}, {\cal K}^{*}]^{T}$ with
the $N \times 1$  matrix  ${\cal Z}$, and $N^{2} \times 1$ matrices ${\cal R}$, ${\cal K}$ given as follows:
\begin{eqnarray}
{\cal Z}_{i} & = & \sum_{klr} V_{iklr} \left [
z^{*(1)}_{k}z^{(0)}_{l} z^{(1)}_{r} + z^{*(1)}_{k}z^{(1)}_{l}
z^{(0)}_{r} + z^{*(0)}_{k}z^{(1)}_{l} z^{(1)}_{r} + 2
\rho^{(001)}_{lk} z^{(1)}_{r} + \kappa^{(1)}_{kl}z^{(1)}_{r}
\right ], \\
{\cal R}_{ij} & = & 2 \sum_{rkl} V_{iklr} \rho^{(1)}_{rj}\rho^{(1)}_{kl} -
V_{rklj} \rho^{(1)}_{kl}\rho^{(1)}_{ir} + \sum_{rkl} V_{irkl}
\kappa^{(1)}_{kl}\kappa^{*(1)}_{rj} + V_{rjkl}
\kappa^{*(1)}_{kl}\kappa^{(1)}_{ir}  \nonumber \\
& + & 2 \sum_{rkl} V_{iklr} z^{*(1)}_{k}z^{(1)}_{l} \rho^{(0)}_{rj} -
V_{rklj} z^{*(1)}_{k}z^{(1)}_{l} \rho^{(0)}_{ir}  \nonumber \\
& + & 2 \sum_{rkl} V_{iklr} \left [ z^{*(1)}_{k}z^{(0)}_{l} +
z^{*(0)}_{k}z^{(1)}_{l} \right ]\rho^{(1)}_{rj} - V_{rklj} \left [
z^{*(1)}_{k}z^{(0)}_{l} + z^{*(0)}_{k}z^{(1)}_{l} \right ] \rho^{(1)}_{ir}
\nonumber \\
& + & \sum_{rkl} V_{rjkl} \left [ z^{*(1)}_{k}z^{*(0)}_{l} +
z^{*(0)}_{k}z^{*(1)}_{l} \right ] \kappa^{(1)}_{ir} - V_{irkl} \left [
z^{(1)}_{k}z^{(0)}_{l} + z^{(0)}_{k}z^{(1)}_{l} \right ]\kappa^{*(1)}_{rj}
\nonumber \\
& + & \sum_{rkl} V_{rjkl} z^{(1)}_{k}z^{(1)}_{l} \kappa^{(0)}_{ir} -
V_{irkl} z^{(1)}_{k}z^{(1)}_{l} \kappa^{*(0)}_{rj}, \\
{\cal K}_{ij} & = & 2 \sum_{rkl} V_{iklr} \kappa^{(1)}_{rj}\rho^{(1)}_{lk} +
V_{rklj} \rho^{(1)}_{kl}\kappa^{(1)}_{ir} + \sum_{rkl} V_{irkl}
\kappa^{(1)}_{kl}\rho^{*(1)}_{rj} + V_{rjkl}
\kappa^{(1)}_{kl}\rho^{(1)}_{ir}
\nonumber \\
& + & 2 \sum_{rkl} V_{iklr} z^{*(1)}_{k}z^{(1)}_{l} \kappa^{(0)}_{rj} +
V_{rklj} z^{(1)}_{k}z^{*(1)}_{l} \kappa^{(0)}_{ir}  \nonumber \\
& + & 2 \sum_{rkl} V_{iklr} \left [ z^{*(1)}_{k}z^{(0)}_{l} +
z^{*(0)}_{k}z^{(1)}_{l} \right ]\kappa^{(1)}_{rj} + V_{rklj} \left [
z^{(1)}_{k}z^{*(0)}_{l} + z^{(0)}_{k}z^{*(1)}_{l} \right ] \kappa^{(1)}_{ir}
\nonumber \\
& + & \sum_{rkl} V_{rjkl} \left [ z^{(1)}_{k}z^{(0)}_{l} +
z^{(0)}_{k}z^{(1)}_{l} \right ] \rho^{(1)}_{ir} + V_{irkl} \left [
z^{(1)}_{k}z^{(0)}_{l} + z^{(0)}_{k}z^{(1)}_{l} \right ]\rho^{*(1)}_{rj}
\nonumber \\
& + & \sum_{rkl} V_{rjkl} z^{(1)}_{k}z^{(1)}_{l} \rho^{(0)}_{ir} + V_{irkl}
z^{(1)}_{k}z^{(1)}_{l} \rho^{*(0)}_{rj} + \sum_{kl} V_{ijkl}
z^{(1)}_{k}z^{(1)}_{l}.
\end{eqnarray}

Casting Eq. (\ref{secondsol}) in the form
\begin{equation}
\vec{\psi}^{(2)}({\bf r},t)=\int \vec{K}^{(2)}(t,t_{1},t_{2},{\bf r},{\bf
r}_{1}
{\bf r}_{2})V_{f}({\bf r}_{1},t_{1})V_{f}({\bf r}_{2},t_{2})d^{3}{\bf
r}_{1}dt_{1}d^{3}{\bf r}_{2}dt_{2}
\end{equation}
involves rewriting $\vec{\Gamma}(t)$ of Eq. (\ref{gamma}) in the position
dependent form:
\begin{eqnarray}
\vec{\Gamma}(t_{1})  & = & \frac{1}{i \hbar} \int_{0}^{t_1} dt_2 \int d{\bf
r}_1
d{\bf r}_2 \left [
\tilde{\Phi}({\bf r}_1)\vec{K}^{(1)}(t_1,t_2,{\bf r}_2) \right ] V_f({\bf
r}_1,t_1)V_f({\bf r}_2,t_2) \\
& & + \int_{0}^{t_1} \int_{0}^{t_1} dt_2 dt_{3} \int d{\bf r}_2 d{\bf r}_{3}
\left [\vec{\Xi}_{K}(t_1-t_2, t_1-t_{3} ; {\bf r}_2,{\bf r}_{3}) \right ]
V_f({\bf r}_2,t_2)V_f({\bf r}_{3},t_{3}) ,
\end{eqnarray}
where $\vec{K}^{(1)}(t_1,t_2,{\bf r}_2)$ is the linear response function for
the
combined variables $z$, $\rho$ and $\kappa$:
\begin{equation}
\vec{K}^{(1)}(t_1-t_2,  {\bf r}_2) = {\cal U}(t_1 -
t_2) \tilde{\Phi}({\bf r}_{2}) \vec{\psi}^{(0)}, \label{compactK1}
\end{equation}
and the $2N(2N+1) \times 1$ column vector $\vec{\Xi}_{K}(t_1-t_2, t_1 -
t_{3} ;
{\bf r}_2,{\bf r}^{\prime}_{2})$ has already been defined above in Eqs. (\ref{XiK}-
\ref{XiKkappa}). $\vec{\Xi}_{K}$ is derived from the vector $\vec{\Xi}(t)$ such that the components
of the linear response vector $\vec{K}^{(1)}(t_1, t_2, {\bf r}_2) $ i.e.
$\vec{K}^{(1)}_{z} (t_1, t_2, {\bf r}_2) $, $\vec{K}^{(1)}_{\rho}(t_1, t_2,
{\bf r}_2) $ and $\vec{K}^{(1)}_{\kappa}(t_1, t_2, {\bf r}_2) $ defined in Eq. (\ref{K_zetc})
replace, respectively, $z^{(1)}(t_1)$, $\rho^{(1)} (t_1) $, and $
\kappa^{(1)} (t_1)$  in $\vec{\Xi}(t)$.

Using these results, the position-dependent time domain
second-order response function for the combined variables $z$,
$\rho$, and $\kappa$ may finally be written as
\begin{equation}
\vec{K}^{(2)}(t,t_{1},t_{2},{\bf r},{\bf r}_{1},{\bf r}_{2})=
\tilde{\Upsilon}(
{\bf r})\left[ \vec{K}_{I}^{(2)}+ \vec{K}_{II}^{(2)} \right],
\label{compactK2}
\end{equation}
where, after the change of variables $t_1 \rightarrow \tau$, $t_2
\rightarrow
t_1$, $t_3 \rightarrow t_2$, $\vec{K}_{I}^{(2)}$ and $\vec{K}_{II}^{(2)}$
are
given  by
\begin{eqnarray}
\vec{K}^{(2)}_{I}(t,t_1,t_2,{\bf r}_1,{\bf r}_2) & = & {\cal U}(t-t_1)
\tilde{\Phi}({\bf r}_1){\cal U}(t_1-t_2) \tilde{\Phi}({\bf
r}_2)\vec{\psi}^{(0)}, \label{KIt} \\
\vec{K}^{(2)}_{II}(t,t_1,t_2,{\bf r}_1,{\bf r}_2) & = & \int_{0}^{t} d \tau
{\cal U}(t-\tau) \vec{\Xi}_{K}(\tau-t_1,\tau - t_2, {\bf r}_1,{\bf r}_2).
\label{KIIt}
\end{eqnarray}

The position-dependent second order response  functions for the
condensate, the non-condensate density and correlation are given
by summing over appropriate indices of the vector $\vec{K}^{(2)}$,
Eq.~(\ref{compactK2}):
\begin{equation}
K^{(2)}_{z}(t,t_1, t_2, {\bf r},{\bf r}_1,{\bf r}_2) = \sum_{i =1}^{N}
\left [\tilde{\Upsilon}({\bf r}) \left ( \vec{K}^{(2)}_{I}(t,t_1,t_2,{\bf
r}_1,
{\bf r}_2) + \vec{K}^{(2)}_{II}(t,t_1,t_2,{\bf r}_1,{\bf r}_2) \right )
\right
]_{i} \label{K2tz1}
\end{equation}
\begin{equation}
K^{(2)}_{\rho}(t, t_1, t_2, {\bf r},{\bf r}_1,{\bf r}_2) = \sum_{i = 2N +
1}^{2N + N^2} \left [\tilde{\Upsilon}({\bf r}) \left (
\vec{K}^{(2)}_{I}(t,t_1,t_2,
{\bf r}_1,{\bf r}_2) + \vec{K}^{(2)}_{II}(t,t_1,t_2,{\bf r}_1,{\bf r}_2)
\right
)
\right ]_{i} \label{K2trho1}
\end{equation}
\begin{equation}
K^{(2)}_{\kappa}(t,t_1, t_2, {\bf r},{\bf r}_1,{\bf r}_2) = \sum_{i = 2N+N^2
+ 1}^{2N+ 2N^2} \left [\tilde{\Upsilon}({\bf r}) \left (
\vec{K}^{(2)}_{I}(t,t_1,t_2,{\bf r}_1,{\bf r}_2) +
\vec{K}^{(2)}_{II}(t,t_1,t_2,{\bf r}
_1,{\bf r}_2) \right ) \right ]_{i}  \label{K2tkappa1}
\end{equation}
where $\vec{K}^{(2)}_{I}$ and $\vec{K}^{(2)}_{II}$ are defined in Eqs.
(\ref{KIt}) -  (\ref{KIIt}), and $\tilde{\Upsilon}$, ${\cal U}$,
$\tilde{\Phi}$
and $\vec{\Xi}_{K}$ are defined in Eqs. (\ref{upsilontilde}), (\ref{U}),
(\ref{phitilde}), and (\ref{XiK}-\ref{XiKkappa}) respectively.

\subsection{Alternative form for the response function}

To discuss the response functions in the frequency domain and to
understand the physical processes involved, it is useful to expand
the response functions in the basis of the eigenvectors
$\vec{\xi}_{\nu}$, of matrix ${\cal L}^{(2)}$ such that
${\cal L}^{(2)} \vec{\xi}_{\nu} = \omega_{\nu} \vec{\xi}_{\nu}$, $\nu = 1,
2,
\ldots 2N(2N+1)$.
We define the Green's function
\begin{equation}
G_{\nu}(t - t') = \exp \left[ -\frac{i}{\hbar }\omega_{\nu}(t-t^{\prime })
\right], \label{G_nu}
\end{equation}
and the expansion coef0ficients $\mu_{\nu}$, $\eta_{\nu}({\bf r})
$, and $\delta_{\nu} ({\bf r})$ such that
\begin{equation}
\vec{\psi}^{(0)} = \sum_{\nu = 1}^{2N(2N+1)} \mu_{\nu} \vec{\xi}_{\nu},
\;\;\;\;  \tilde{\Phi}({\bf r}) \vec{\xi}_{\nu'} = \sum_{\nu = 1}^{2N(2N+1)}
\eta_{\nu}({\bf r}) \vec{\xi}_{\nu}, \;\;\;\; {\rm and} \;\;\;\;
\tilde{\Upsilon} ({\bf r})  \vec{\xi}_{\nu} = \sum_{\nu = 1}^{2N(2N+1)}
\delta_{\nu} ({\bf r}) \vec{\xi}_{\nu}. \label{munudelta}
\end{equation}

In the basis of these eigenstates, $\vec{\xi}_{\nu}$, the result is Eqs.
(\ref{K2tz1})-(\ref{K2tkappa1}) but with
\begin{eqnarray}
\vec{K}^{(2)}_{I}(t,t_1,t_2,{\bf r}_1,{\bf r}_2) & = & \sum_{\nu =
1}^{2N(2N+1)}
{\cal K}^{(2)}_{I, \nu}(t,t_1,t_2,{\bf r}_1,{\bf r}_2) \vec{\xi}_{\nu} ,\\
\vec{K}^{(2)}_{II}(t,t_1,t_2,{\bf r}_1,{\bf r}_2) & = & \sum_{\nu  =
1}^{2N(2N+1)} {\cal K}^{(2)}_{II, \nu}(t,t_1,t_2,{\bf r}_1,{\bf r}_2)
\vec{\xi}_{\nu} ,
\end{eqnarray}
where
\begin{equation}
{\cal K}^{(2)}_{I, \nu}(t,t_1,t_2,{\bf r}_1,{\bf r}_2)   =  \sum_{\nu',
\nu'' =
1}^{2N(2N+1)} \eta_{\nu}({\bf r}_1) \eta_{\nu'}({\bf r}_2) G_{\nu}(t-t_1)
\mu_{\nu''} G_{\nu'}(t_1-t_2),
\end{equation}
and
\begin{equation}
{\cal K}^{(2)}_{II, \nu}(t,t_1,t_2,{\bf r}_1,{\bf r}_2)  =  \sum_{\nu',
\nu'' =
1}^{2N(2N+1)} \int_{0}^{t} d \tau G_{\nu}(t-\tau) {\cal F} \left [
\eta_{\nu'}({\bf r}_1) \eta_{\nu''}({\bf r}_2) \mu_{\nu'} \mu_{\nu''}
G_{\nu'}(\tau-t_1)G_{\nu''}(\tau - t_2) \right ],
\end{equation}
where ${\cal F}$ is the function given according to the expression for
$\vec{\Xi}_{K}$ but written in terms of $\mu_{\nu}$, $\eta_{\nu}$, and
$G_{\nu}$ with  $G_{\nu}(t - t')$, $\mu_{\nu}$, and $\eta_{\nu}$ as defined
 in Eqs. (\ref{G_nu}-\ref{munudelta}).

\section{Second order susceptibilities} \label{chi2}

\subsection{Final expression}
The final result that we use for our numerical calculation for the
condensate,
non-condensate density
and the non-condensate correlations are:
\begin{equation}
K^{(2)}_{z}(-\Omega_{1} - \Omega_{2}; \Omega_{1}, \Omega_{2}, {\bf r},{\bf
r}%
_1,{\bf r}_2) = \sum_{i =1}^{N}    \left [ \tilde {\Upsilon}({\bf r}) \left
(
\vec{K}^{(2)}_{I}(\Omega_{1}, \Omega_{2}, {\bf r}_1,{\bf r}_2) +
\vec{K}^{(2)}_{II}(\Omega_{1}, \Omega_{2}, {\bf r%
}_1,{\bf r}_2) \right ) \right ]_{i}  \label{K2wz}
\end{equation}
\begin{equation}
K^{(2)}_{\rho}(-\Omega_{1} - \Omega_{2}; \Omega_{1}, \Omega_{2}, {\bf r},%
{\bf r}_1,{\bf r}_2) = \sum_{i = 2N + 1}^{2N + N^2} \left [ \tilde
{\Upsilon}({\bf r}) \left (
\vec{K}^{(2)}_{I}(\Omega_{1}, \Omega_{2},  {\bf r%
}_1,{\bf r}_2) + \vec{K}^{(2)}_{II}(\Omega_{1},
\Omega_{2}, {\bf r}_1,{\bf r}_2) \right ) \right ]_{i} \label{K2wrho}
\end{equation}
\begin{equation}
K^{(2)}_{\kappa}(-\Omega_{1} - \Omega_{2}; \Omega_{1}, \Omega_{2}, {\bf r},%
{\bf r}_1,{\bf r}_2) = \sum_{i = 2N+N^2 + 1}^{2N+ 2N^2} \left [ \tilde
{\Upsilon}({\bf r}) \left (
\vec{K}^{(2)}_{I}(\Omega_{1}, \Omega_{2},  {\bf r%
}_1,{\bf r}_2) + \vec{K}^{(2)}_{II}(\Omega_{1},
\Omega_{2},  {\bf r}_1,{\bf r}_2) \right ) \right ]_{i}. \label{K2wkappa}
\end{equation}
where
\begin{equation}
\vec{K}^{(2)}_{I}(\Omega_{1}, \Omega_{2}, {\bf r},{\bf r}_1, {\bf r}_2)  =
-
\frac{1}{4 \pi^2} \tilde{\Upsilon}({\bf r}) {\cal U}(\Omega_1 + \Omega_2)
\tilde{\Phi}({\bf r}_1) {\cal U}(\Omega_2) \tilde{\Phi}({\bf
r}_2)\vec{\psi}^{(0)} ,  \label{K2Iomega1}
\end{equation}
and
\begin{equation}
\vec{K}^{(2)}_{II}(\Omega_{1}, \Omega_{2}, {\bf r},{\bf %
r}_1, {\bf r}_2) = -\frac{1}{8 \pi^3 i} \tilde {\Upsilon}({\bf r}) {\cal U}%
(\Omega_{1} + \Omega_{2}) \vec{\Xi}_{K}(\Omega_{1}, \Omega_{2}, {\bf r}%
_1,{\bf r}_2).  \label{K2IIomega1}
\end{equation}
Here, $\tilde {\Upsilon}({\bf r})$ and $\tilde{\Phi}({\bf r})$ are as
defined in
Eqs. (\ref{upsilontilde}) and (\ref{phitilde}) and
\begin{equation}
{\cal U}(\omega) \equiv \frac{1}{\omega - {\cal L}^{(2)} + i \epsilon}.
\label{Uomega}
\end{equation}
In addition, the vector $\vec{\Xi}_{K}(\Omega_{1}, \Omega_{2}, {\bf r}_1,%
{\bf r}_2)$ of  Eq. (\ref{K2IIomega1}) is may be written as
\begin{equation}
\vec{\Xi}_{K}(\Omega_{1}, \Omega_{2}, {\bf r}_1,%
{\bf r}_2) = [\tilde{{\cal Z}}_{K}, \tilde{{\cal Z}}^{*}_{K}, \tilde{{\cal
R}}_{K}, \tilde{{\cal K}}_{K}, \tilde{{\cal R}}^{*}_{K}, \tilde{{\cal
K}}^{*}_{K}]^{T} \label{XiKw}
\end{equation}
with the $N \times 1$ matrix $\tilde{{\cal Z}}_{K}$ and $N^{2} \times 1$ matrices $\tilde{{\cal R}}_{K}$, $\tilde{{\cal K}}_{K}$ given as follows:
\begin{eqnarray}
\left [ \tilde{{\cal Z}}_{K} \right ]_{i} & = & \sum_{klr} V_{iklr} \left [
[\vec{K}_z^{(1)*}(\Omega_1, {\bf r}_{1})]_{k} z^{(0)}_{l}
[\vec{K}_z^{(1)} (\Omega_2, {\bf r}_{2})]_{r} + [\vec{K}_z^{(1)*}(\Omega _1,
{\bf r}_{1})]_{k}[\vec{K}_z^{(1)}(\Omega_2, {\bf r}_{2})]_{l} z^{(0)}_{r}
\right. \nonumber \\
& + & z^{*(0)}_{k}[\vec{K}_z^{(1)}(\Omega_1, {\bf r}_{1})]_{l}
[\vec{K}_z^{(1)}(\Omega_2, {\bf r}_{2})]_{r} + 2
[\vec{K}_\rho^{(1)}(\Omega_1,
{\bf r}_{1})]_{lk} [\vec{K}_z^{(1)}(\Omega_2, {\bf r}_{2})]_{r}   \nonumber
\\
& + & \left. [\vec{K}_\kappa^{(1)}(\Omega_1, {\bf
r}_{1})]_{kl}[\vec{K}_z^{(1)}(\Omega_2, {\bf r}_{2})]_{r}
\right ]  \label{XiKwz} \\
\left [ \tilde{{\cal R}}_{K} \right ]_{ij}  & = & 2 \sum_{rkl} V_{iklr}
[\vec{K}_\rho^{(1)}(\Omega_1, {\bf
r}_{1})]_{rj}[\vec{K}_\rho^{(1)}(\Omega_2,
{\bf r}_{2})]_{kl} -
V_{rklj} [\vec{K}_\rho^{(1)}(\Omega_1, {\bf
r}_{1})]_{kl}[\vec{K}_\rho^{(1)}(\Omega_2, {\bf r}_{2})]_{ir}  \nonumber  \\
& + & \sum_{rkl} V_{irkl} [\vec{K}_\kappa^{(1)}(\Omega_1, {\bf
r}_{1})]_{kl}[\vec{K}0_\kappa^{*(1)}(\Omega_2, {\bf r}_{2})]_{rj}
+ V_{rjkl} [\vec{K}_\kappa^{*(1)}(\Omega_1, {\bf
r}_{1})]_{kl}[\vec{K}_\kappa^{(1)}(\Omega_2, {\bf r}_{2})]_{ir}
\nonumber
\\
& + & 2 \sum_{rkl} V_{iklr} [\vec{K}_z^{*(1)}(\Omega_1, {\bf
r}_{1})]_{k}[\vec{K}_z^{(1)}(\Omega_2, {\bf r}_{2})]_{l} \rho^{(0)}_{rj} -
V_{rklj} [\vec{K}_z^{*(1)}(\Omega_1, {\bf
r}_{1})]_{k}[\vec{K}_z^{(1)}(\Omega_2,
{\bf r}_{2})]_{l} \rho^{(0)}_{ir}  \nonumber \\
& + & 2 \sum_{rkl} V_{iklr} \left [ [\vec{K}_z^{*(1)}(\Omega_1, {\bf
r}_{1})]_{k}z^{(0)}_{l} +
z^{*(0)}_{k}[\vec{K}_z^{(1)}(\Omega_1, {\bf r}_{1})]_{l} \right ]
[\vec{K}_\rho^{(1)}(\Omega_2, {\bf r}_{2})]_{rj} \nonumber \\
& - & V_{rklj} \left [
[\vec{K}_z^{*(1)}(\Omega_1, {\bf r}_{1})]_{k}z^{(0)}_{l} +
z^{*(0)}_{k}[\vec{K}_z^{(1)}(\Omega_1, {\bf r}_{1})]_{l} \right ]
[\vec{K}_\rho^{(1)}(\Omega_2, {\bf r}_{2})]_{ir}
\nonumber \\
& + & \sum_{rkl} V_{rjkl} \left [ [\vec{K}_z^{(1)*}(\Omega_1, {\bf
r}_{1})]_{k}z^{*(0)}_{l} +
z^{*(0)}_{k}[\vec{K}_z^{*(1)}(\Omega_1, {\bf r}_{1})]_{l} \right ]
[\vec{K}_\kappa^{(1)}(\Omega_2, {\bf r}_{2})]_{ir} \nonumber \\
& - & V_{irkl} \left [
[\vec{K}_z^{(1)}(\Omega_1, {\bf r}_{1})]_{k}z^{(0)}_{l} +
z^{(0)}_{k}[\vec{K}_z^{(1)}(\Omega_1, {\bf r}_{1})]_{l} \right ]
[\vec{K}_\kappa^{*(1)}(\Omega_2, {\bf r}_{2})]_{rj}
\nonumber \\
& + & \sum_{rkl} V_{rjkl} [\vec{K}_z^{(1)}(\Omega_1, {\bf
r}_{1})]_{k}[\vec{K}_z^{(1)}(\Omega_2, {\bf r}_{2})]_{l} \kappa^{(0)}_{ir} -
V_{irkl} [\vec{K}_z^{(1)}(\Omega_1, {\bf
r}_{1})]_{k}[\vec{K}_z^{(1)}(\Omega_2,
{\bf r}_{2})]_{l} \kappa^{*(0)}_{rj} \label{XiKwrho} \\
\left [ \tilde{{\cal K}}_{K} \right ]_{ij} & = & 2 \sum_{rkl} V_{iklr}
[\vec{K}_\kappa^{(1)}(\Omega_1, {\bf
r}_{1})]_{rj}[\vec{K}_\rho^{(1)}(\Omega_2,
{\bf r}_{2})]_{lk} +
V_{rklj} [\vec{K}_\rho^{(1)}(\Omega_1, {\bf
r}_{1})]_{kl}[\vec{K}_\kappa^{(1)}(\Omega_2, {\bf r}_{2})]_{ir} \nonumber \\
& + & \sum_{rkl} V_{irkl}
[\vec{K}_\kappa^{(1)}(\Omega_1, {\bf
r}_{1})]_{kl}[\vec{K}_\rho^{(1)*}(\Omega_2,
{\bf r}_{2})]_{rj} + V_{rjkl} [\vec{K}_\kappa^{(1)}(\Omega_1, {\bf
r}_{1})]_{kl}[\vec{K}_\rho^{(1)}(\Omega_2, {\bf r}_{2})]_{ir}
\nonumber \\
& + & 2 \sum_{rkl} V_{iklr} [\vec{K}_z^{*(1)}(\Omega_1, {\bf
r}_{1})]_{k}[\vec{K}_z^{(1)}(\Omega_2, {\bf r}_{2})]_{l} \kappa^{(0)}_{rj} +
V_{rklj} [\vec{K}_z^{(1)}(\Omega_1, {\bf
r}_{1})]_{k}[\vec{K}_z^{*(1)}(\Omega_2,
{\bf r}_{2})]_{l} \kappa^{(0)}_{ir}  \nonumber \\
& + & 2 \sum_{rkl} V_{iklr} \left [ [\vec{K}_z^{(1)*}(\Omega_1,
{\bf r}_{1})]_{k}z^{0(0)}_{l} +
z^{*(0)}_{k}[\vec{K}_z^{(1)}(\Omega_1, {\bf r}_{1})]_{l} \right
][\vec{K}_\kappa^{(2)}(\Omega_1, {\bf r}_{2})]_{rj} \nonumber \\
& + & V_{rklj} \left [
[\vec{K}_z^{(1)}(\Omega_1, {\bf r}_{1})]_{k}z^{*(0)}_{l} +
z^{(0)}_{k}[\vec{K}_z^{*(1)}(\Omega_1, {\bf r}_{1})]_{l} \right ]
[\vec{K}_\kappa^{(1)}(\Omega_2, {\bf r}_{2})]_{ir}
\nonumber \\
& + & \sum_{rkl} V_{rjkl} \left [ [\vec{K}_z^{(1)}(\Omega_1, {\bf
r}_{1})]_{k}z^{(0)}_{l} +
z^{(0)}_{k}[\vec{K}_z^{(1)}(\Omega_1, {\bf r}_{1})]_{l} \right ]
[\vec{K}_\rho^{(1)}(\Omega_2, {\bf r}_{2})]_{ir}  \nonumber \\
& + & V_{irkl} \left [
[\vec{K}_z^{(1)}(\Omega_1, {\bf r}_{1})]_{k}z^{(0)}_{l} +
z^{(0)}_{k}[\vec{K}_z^{(1)}(\Omega_1, {\bf r}_{1})]_{l} \right ]
[\vec{K}_\rho^{*(1)}(\Omega_2, {\bf r}_{2})]_{rj}
\nonumber \\
& + & \sum_{rkl} V_{rjkl} [\vec{K}_z^{(1)}(\Omega_1, {\bf
r}_{1})]_{k}[\vec{K}_z^{(1)}(\Omega_2, {\bf r}_{2})]_{l} \rho^{(0)}_{ir}
+ V_{irkl}
[\vec{K}_z^{(1)}(\Omega_1, {\bf r}_{1})]_{k}[\vec{K}_z^{(1)}(\Omega_2, {\bf
r}_{2})]_{l} \rho^{*(0)}_{rj} \nonumber \\
& + & \sum_{kl} V_{ijkl}
[\vec{K}_z^{(1)}(\Omega_1, {\bf r}_{1})]_{k}[\vec{K}_z^{(1)}(\Omega_2, {\bf
r}_{2})]_{l}. \label{XiKwkappa}
\end{eqnarray}0
Similar to Eq. (\ref{K_zetc}) above,  the
quantities
$\vec{K}^{(1)}_{z}(\Omega_i, {\bf r}_i)$, $\vec{K}^{(1)}_{\rho}(\Omega_i, {\bf
r}_i)$ and
$\vec{K}^{(1)}_{\kappa}(\Omega_i, {\bf r}_i)$ used in Eqs. (\ref{XiKwz}-\ref{XiKwkappa}) are defined as
\begin{equation}
\vec{K}^{(1)}_{\alpha}(\Omega_i,{\bf r}_i) =  {\cal U}^{(\alpha)}(\Omega_i)
\tilde{\Phi}({\bf r}_i) \vec{\psi}^{(0)} , \label{K_zwetc}
\end{equation}
where $\alpha = z, \rho, \kappa$, and $i= 1,2$. ${\cal U}^{(\alpha)}(\Omega_i)$ are the
submatrices of ${\cal U}(\Omega_i)$ defined in Eq. (\ref{Uomega}) such that
\begin{equation}
{\cal U}(\Omega_i) = \left [ {\cal U}^{(z)}(\Omega_i), {\cal U}^{(z)*}(\Omega_i),
{\cal U}^{(\rho)}(\Omega_i), {\cal U}^{(\kappa)}(\Omega_i), {\cal
U}^{(\rho)*}(\Omega_i),
{\cal U}^{(\kappa)*}(\Omega_i) \right ]^{T}
\end{equation}
where ${\cal U}^{(z)}(\Omega_i)$ is an $N \times 2N(2N + 1)$ submatrix while
$%
{\cal U}^{(\gamma)}(\Omega_i)$, $\gamma = \rho, \kappa, \rho^{*},
\kappa^{*}$ is an $N^2 \times 2N(2N+1)$ submatrix such that the
submatrix ${\cal U}^{(z)}(\Omega_i)$ is stacked on top of
submatrix ${\cal U}^{(z)*}(\Omega_i)$ which, in turn, is stacked
on top of submatrices ${\cal U}^{(\gamma)}(\Omega_i)$. It is to be
noted that, as with the time domain example discussed above,
$\vec{K}^{(1)}_{\alpha}(\Omega_i, {\bf r}_i)$ as defined here are
$N \times 1$ and $N^2 \times 1$ {\em vectors}, not scalar
quantities obtained by integrating the scalar function
$K^{(1)}_{\alpha}(\Omega_i, {\bf r}, {\bf r}_i)$ over ${\bf r}$.

\subsection{Derivation}
The second order response function in frequency is given by the Fourier
Transform of the time domain counterpart:
\begin{eqnarray}
\vec{K}^{(2)}( \Omega, \Omega_{1}, \Omega_{2}, {\bf r},{\bf r}_1, {\bf r}_2)
& = & \int_{0}^{\infty} dt \/ dt_{1} \tilde{\Upsilon}({\bf
r})\left[ \vec{K}_{I}^{(2)} + \vec{K}_{II}^{(2)} \right] \exp \left ( i
\Omega t
+ i
\Omega_{1} t_{1} + i \Omega_{2} t_{2} \right ) \\
& = & \tilde{\Upsilon}({\bf r}) \left[ \vec{K}^{(2)}_{I}( \Omega,
\Omega_{1},
\Omega_{2}, {\bf r}_1, {\bf r}_2) + \vec{K}^{(2)}_{II}( \Omega, \Omega_{1},
\Omega_{2}, {\bf r}_1, {\bf r}_2) \right], \label{freqdom_K2}
\end{eqnarray}
where $\vec{K}^{(2)}_{I}( \Omega, \Omega_{1}, \Omega_{2}, {\bf r}_1, {\bf
r}_2)$
and $\vec{K}^{(2)}_{II}( \Omega, \Omega_{1}, \Omega_{2}, {\bf r}_1, {\bf
r}_2)$
are the Fourier transforms of the time domain expressions Eq. (\ref{KIt})-
(\ref{KIIt}).

Using the fact that the matrices ${\cal U}(t-t_1)$ are the Green's
functions with an implicit Heaviside step function in time i.e.
${\cal U}(t-t_1) \equiv \theta(t-t_1) {\cal U}(t-t_1)$ such that
\begin{eqnarray}
\theta(t) {\cal U}(t) & = & - \frac{1}{2 \pi i} \int_{-\infty}^{\infty} d
\omega
\frac{1}{\omega - {\cal L}^{(2)} + i \epsilon} \exp(- i \omega t) \\
& = & \int_{-\infty}^{\infty} d \omega
{\cal U}(\omega) \exp(- i \omega t),  \label{FT}
\end{eqnarray}
we have
\begin{eqnarray}
\vec{K}^{(2)}_{I}( \Omega, \Omega_{1}, \Omega_{2}, {\bf r},{\bf r}_1, {\bf
r}_2)
& = & -\frac{1}{4 \pi^2} \int_{-\infty}^{\infty} d \omega d
\omega^{\prime}\; \; \tilde{\Upsilon}({\bf r}) {\cal U}(\omega)
\tilde{\Phi}({\bf r}_1){\cal U}(\omega^{\prime}) \tilde{\Phi}({\bf
r}_2)\vec{\psi}^{(0)}
\delta(\Omega - \omega)  \nonumber \\
& & \times \delta(\Omega_{1} + \omega - \omega^{\prime}) \delta(\Omega_{2} +
\omega^{\prime}) .
\end{eqnarray}
This implies that $\omega^{\prime}= - \Omega_{2}$, $\omega = -\Omega_{1} -
\Omega_{2}$, and $\Omega = - \Omega_{1} - \Omega_{2}$:
\begin{equation}
\vec{K}^{(2)}_{I}(-\Omega_{1}-\Omega_{2}; \Omega_{1}, \Omega_{2}, {\bf
r},{\bf
r}_1, {\bf r}_2)  =  -\frac{1}{4 \pi^2} \tilde{\Upsilon}({\bf r}) {\cal
U}(\Omega_1 + \Omega_2)
\tilde{\Phi}({\bf r}_1) {\cal U}(\Omega_2) \tilde{\Phi}({\bf
r}_2)\vec{\psi}^{(0)}.  \label{K2Iomega}
\end{equation}
In addition we have,
\begin{eqnarray}
\vec{K}^{(2)}_{II}( \Omega, \Omega_{1}, \Omega_{2}, {\bf r},{\bf r}_1, {\bf
r}_2) & = &
-\frac{1}{8 \pi^3 i} \int_{-\infty}^{\infty} d \omega d
\omega^{\prime}d \omega^{\prime\prime}\;\;\; \tilde {\Upsilon}({\bf r})
{\cal U}(\omega) \vec{\Xi}_{K}(\omega^{\prime}, \omega^{\prime\prime},
{\bf r}_1,{\bf r}_2)  \nonumber \\
& \times &  \delta(\omega - \omega ^{\prime}- \omega
^{\prime\prime}) \delta(\Omega - \omega) \delta(\Omega_{1} + \omega^{\prime})
\delta (\Omega_{2} + \omega ^{\prime\prime}) \label{K2exp} .
\end{eqnarray}
We are able to write the Fourier Transform for $\vec{\Xi}_{K}(t)$ in Eq.
(\ref{K2exp}) since the function $\vec{\Xi}_{K}(t)$ is made up  of terms
which
are simply products of two Green's functions at different times.  Eq.
(\ref{K2exp}) implies $\omega^{\prime}= -\Omega_{1}$,
$\omega^{\prime\prime}=
-\Omega_{2}$, $\omega = \omega^{\prime}+ \omega^{\prime\prime}= -\Omega_{1}
- \Omega_{2}$, $\Omega = \omega$ so that
\begin{equation}
\vec{K}^{(2)}_{II}( -\Omega_{1} - \Omega_{2}; \Omega_{1}, \Omega_{2}, {\bf
r},{\bf r}_1, {\bf r}_2) = -\frac{1}{8 \pi^3 i} \tilde {\Upsilon}({\bf r}) {\cal U}(\Omega_{1}
+ \Omega_{2}) \vec{\Xi}_{K}(\Omega_{1}, \Omega_{2}, {\bf r}_1,{\bf r}_2),   \label{K2IIw}
\end{equation}
where $\vec{\Xi}_{K}(\Omega_{1}, \Omega_{2}, {\bf r}_1,{\bf r}_2)$ is as already given in Eq. (\ref{XiKwkappa}).

As for the time domain calculations, the susceptibilities for the
condensate,
non-condensate density
and the non-condensate correlations are obtained by summing over the
appropriate indices:
\begin{equation}
K^{(2)}_{z}(-\Omega_{1} - \Omega_{2}; \Omega_{1}, \Omega_{2}, {\bf r},{\bf
r}_1,{\bf r}_2) = \sum_{i =1}^{N}    \left [ \tilde {\Upsilon}({\bf r}) \left
(
\vec{K}^{(2)}_{I}(\Omega_{1}, \Omega_{2}, {\bf r}_1,{\bf r}_2) +
\vec{K}^{(2)}_{II}(\Omega_{1}, \Omega_{2}, {\bf r}_1,{\bf r}_2) \right ) \right ]_{i}  \label{K2wz1}
\end{equation}
\begin{equation}
K^{(2)}_{\rho}(-\Omega_{1} - \Omega_{2}; \Omega_{1}, \Omega_{2}, {\bf r},
{\bf r}_1,{\bf r}_2) = \sum_{i = 2N + 1}^{2N + N^2} \left [ \tilde
{\Upsilon}({\bf r}) \left (
\vec{K}^{(2)}_{I}(\Omega_{1}, \Omega_{2},  {\bf r}_1,{\bf r}_2) + \vec{K}^{(2)}_{II}(\Omega_{1},
\Omega_{2}, {\bf r}_1,{\bf r}_2) \right ) \right ]_{i} \label{K2wrho1}
\end{equation}
\begin{equation}
K^{(2)}_{\kappa}(-\Omega_{1} - \Omega_{2}; \Omega_{1}, \Omega_{2}, {\bf r},%
{\bf r}_1,{\bf r}_2) = \sum_{i = 2N+N^2 + 1}^{2N+ 2N^2} \left [ \tilde
{\Upsilon}({\bf r}) \left (
\vec{K}^{(2)}_{I}(\Omega_{1}, \Omega_{2},  {\bf r}_1,{\bf r}_2) + \vec{K}^{(2)}_{II}(\Omega_{1},
\Omega_{2},  {\bf r}_1,{\bf r}_2) \right ) \right ]_{i} \label{K2wkappa1}
\end{equation}

\subsection{Alternative form for the susceptibility}
As before, expanding in the eigenstate basis $\vec{\xi}_{\nu}$ which we
introduced in Eq. (\ref{munudelta}), we may write the susceptibility in a
more
useful form.
The result is Eqs. (\ref{K2wz1}) - (\ref{K2wkappa1})  but with
\begin{eqnarray}
\vec{K}^{(2)}_{I}(-\Omega_{1}-\Omega_{2}; \Omega_{1}, \Omega_{2},{\bf
r}_1,{\bf
r}_2) & = & \sum_{\nu = 1}^{2N(2N+1)} {\cal K}^{(2)}_{I, \nu}(-\Omega_{1}-
\Omega_{2}; \Omega_{1}, \Omega_{2}, {\bf r}_1,{\bf r}_2) \vec{\xi}_{\nu} ,\\
\vec{K}^{(2)}_{II}(-\Omega_{1}-\Omega_{2}; \Omega_{1}, \Omega_{2},{\bf
r}_1,{\bf
r}_2) & = & \sum_{\nu  = 1}^{2N(2N+1)} {\cal K}^{(2)}_{II, \nu}(-\Omega_{1}-
\Omega_{2}; \Omega_{1}, \Omega_{2}, {\bf r}_1,{\bf r}_2) \vec{\xi}_{\nu} ,
\end{eqnarray}
where
\begin{equation}
{\cal K}^{(2)}_{I, \nu}(-\Omega_{1}-\Omega_{2}; \Omega_{1}, \Omega_{2}, {\bf
r}_1, {\bf r}_2) = -\frac{1}{4 \pi^2} \sum_{\nu', \nu'' = 1}^{2N(2N+1)}
\frac{
\eta_{\nu}({\bf r}_1)\eta_{\nu'}({\bf r}_2) \mu_{\nu''}}{(\Omega_{1} +
\Omega_{2} -\omega_{\nu}   + i \epsilon)
(\Omega_{2} - \omega_{\nu'}  + i \epsilon)}.
\end{equation}
Since no additional information is gained by listing all the terms in ${\cal K}^{(2)}_{II, \nu}(-\Omega_{1}-\Omega_{2}; \Omega_{1}, \Omega_{2}, {\bf r}_1,{\bf r}_2)$, we simply note that, in the eigenstate basis, the typical term in ${\cal K}^{(2)}_{II, \nu}$ has the  structure:
\begin{equation}
  \sum_{\nu' \nu'' = 1}^{2N(2N+1)}  \frac{\eta_{\nu}({\bf
r}_1)\eta_{\nu'}({\bf
r}_2) \mu_{\nu''}}{(\Omega_{1} + \Omega_{2} -\omega_{\nu}  + i
\epsilon)(\Omega_{1} - \omega_{\nu}  + i \epsilon)(\Omega_{2} -\omega_{\nu'}
  + i \epsilon)},
\end{equation}
as to be expected from Eqs. (\ref{XiKwkappa}) and (\ref{K2IIw}).
$\mu_{\nu}$, and $\eta_{\nu}$ are as given in Eq.(\ref{munudelta}).


\newpage
\vspace{2cm} \noindent Fig. 1: Natural log of linear
susceptibility $K^{(1)}(\Omega, {\bf r}, {\bf r}_1)$  at ${\bf r}
= {\bf r}_1 = 0$ vs. frequency. Top three panels --  zero
temperature, bottom three panels -- finite temperature $10
\hbar\omega/k$.  The frequency $\Omega_1$ are given in units of
the trap frequency.

\vspace{2cm} \noindent Fig. 2:  $|K^{(2)}(t,t_1,t_2, {\bf r}, {\bf
r}_1 {\bf r}_2)|$ i.e. the absolute value of the second order
response functions in the time domain with the time $t_2$ fixed at
$t_2 = 0$. (a)  ${\bf r}_2 = 0$; (b) ${\bf r}_2 = -5$.   The plots
are for zero temperature condensate at the short, intermediate and
long times $t$ and $t_1$ written at the top of each column of
figures.  The top, middle, and bottom rows give the response
function for the condensate, non-condensate density, and
non-condensate correlation respectively. The diameter of the
dashed circle  represent the spatial extent of the trapped BEC.
The position  $x$ and $x'$ are given in harmonic oscillator length
units.  \label{tT0r49_65}

\vspace{2cm} \noindent Fig. 3: Same as in Fig. 2, but at finite
temperature of $10 \hbar\omega/k$. The position  $x$ and $x'$ are
given in harmonic oscillator length units. \label{tT10r49_65}

\vspace{2cm} \noindent Fig. 4: $|K^{(2)}(\Omega,\Omega_1,\Omega_2,
{\bf r}, {\bf r}_1 {\bf r}_2)|$ i.e. the absolute value of the
second order response functions in the frequency domain with the
variable ${\bf r}_2$ set at (a) ${\bf r}_2 = 0$ and (b) ${\bf r}_2
= -5$. The plots are for zero temperature condensate at the
frequencies $\Omega_1$ and $\Omega_2$ given at the top of each
column.  Denoting ``off-resonant'' when the frequency dos not
match an eigen value of ${\cal L}^{(2)}$ and ``on-resonance'' when
a frequency matches an eigen value, these frequencies are chosen
such  that $\Omega_1$, $\Omega_2$, and $\Omega_1 + \Omega_2$ are
off-resonant ($\Omega_1 = 2.23$, $\Omega_2 = 1.55$); both
$\Omega_1$ and $\Omega_2$ are on-resonance while $\Omega_1 +
\Omega_2$ is off-resonant ($\Omega_1 = 2.2$, $\Omega_2 = 1.5$);
and finally frequencies chosen so that $\Omega_1 + \Omega_2$ is
on-resonance ($\Omega_1 = 0.7$, $\Omega_2 = 1.5$). The top,
middle, and bottom rows give the response function for the
condensate, non-condensate density, and non-condensate correlation
respectively. The diameter of the dashed circle  represent the
spatial extent of the trapped BEC. The position  $x$ and $x'$ are
given in harmonic oscillator length units.  \label{fT0r49_65}

\vspace{2cm} \noindent Fig. 5:  Same as in Fig. 4, but at finite
temperature of $10 \hbar\omega/k$.  At finite temperature, the
resonant frequencies are shifted from the zero temperature
counterpart  so that the actual frequency combinations used are
the following: $\Omega_1$, $\Omega_2$, and $\Omega_1 + \Omega_2$
off-resonant ($\Omega_1 = 2.45$, $\Omega_2 = 1.6$); both
$\Omega_1$ and $\Omega_2$ on-resonance with $\Omega_1 + \Omega_2$
off- resonant ($\Omega_1 = 2.43$, $\Omega_2 = 1.5$); and finally
$\Omega_1 + \Omega_2$  on-resonance ($\Omega_1 = 0.92$, $\Omega_2
= 1.5$). The position  $x$ and $x'$ are given in harmonic
oscillator length units.   \label{fT10r49_65}

\vspace{2cm} \noindent Fig. 6:  Left column:
$|K^{(2)}(\Omega_1,\Omega_2)|$ i.e. the absolute value of the
second order response functions in the frequency domain as a
function of $\Omega_1$ and $\Omega_2$ with the position variables
set at ${\bf r} = {\bf r}_1 = {\bf r}_2 = 0$. There are three
large peaks that completely dominates the plot so that the
remaining peaks are not represented in the plot. Center column:
Scaled  $|K^{(2)}(\Omega_1,\Omega_2)|$. The three largest peaks
shown in the left column were scaled down to the same magnitude as
other peaks in the plot.  Right column: ${\rm
log}|K^{(2)}(\Omega_1,\Omega_2)|$ i.e. logarithm of the response
functions presented in the left column to help visualize the large
variation in the magnitude of $|K^{(2)}(\Omega_1,\Omega_2)|$. The
top, middle, and bottom rows give the response function for the
condensate, non-condensate density, and non-condensate correlation
respectively. The frequency $\Omega_1$ and $\Omega_2$ are given in
units of the trap frequency.  \label{K2wwT0}

\vspace{2cm} \noindent Fig. 7: Same as in Fig. 6, but at finite
temperature of $10 \hbar\omega/k$. The frequency $\Omega_1$ and
$\Omega_2$ are given in units of the trap frequency.
\label{K2wwT10}


\newpage


\begin{thebibliography}{99}

\bibitem{choi} S. Choi, V. Chernyak, S. Mukamel, \emph{Phys. Rev.
A} {\bf 67}, 043602 (2003).


\bibitem{Band} M. Trippenbach, Y. B. Band and P. S. Julienne  Phys. Rev. A
{\bf 62} 023608 (2000)


\bibitem{CastinDum} Y. Castin, and R. Dum Phys. Rev. A  {\bf 57}  3008 (1998); Y. Castin and R. Dum Phys. Rev. Lett. {\bf 79} 3553 (1997)


\bibitem{Ring} P. Ring and P. Schuck, {\em The Nuclear Many-body problem} (Springer-Verlag, New York, 1980)


\bibitem{Blaizot_Ripka}  J.-P. Blaizot and G. Ripka, {\em Quantum Theory of
Finite Systems} (MIT Press, Cambridge MA, 1986)

\bibitem{Griffin}  A. Griffin, Phys. Rev. B {\bf 53}, 9341 (1996)

\bibitem{Prou2}  N.P. Proukakis and K. Burnett, J. Res. Natl. Inst. Stand.
Technol. {\bf 101}, 457 (1996)


\bibitem{Baym} L. P. Kandanoff and G. Baym {\it Quantum Statistical Mechanics} (W. A. Benjamin  Inc. New York, 1962)

\bibitem{Martin} P.C. Hohenberg and P. C. Martin Ann. Phys.(N.Y.) {\bf 34} 291 (1965) [reprinted {\bf 281} 636 (2000)]

\bibitem{Gardiner} C. W. Gardiner, and P. Zoller Phys. Rev. A  {\bf 55}  2902 (1997);
D. Jaksch, C. W. Gardiner, and P. Zoller Phys. Rev. A  {\bf 56}
575 (1997); C. W. Gardiner, and P. Zoller Phys. Rev. A  {\bf 58}
536 (1998); D. Jaksch, C. W. Gardiner, K. M. Gheri, and P. Zoller
Phys. Rev. A  {\bf 58}  1450 (1998); C. W. Gardiner and P. Zoller
Phys. Rev. A {\bf 61} 033601 (2000)

\bibitem{ZNG} E. Zaremba, T. Nikuni, A. Griffin, J. Low Temp. Phys. {\bf 116} 277 (1999)

\bibitem{Walser} R. Walser, J. Williams, J. Cooper, and M. Holland Phys. Rev. A  {\bf 59}  3878 (1999); R. Walser, J. Cooper, and M. Holland Phys. Rev. A  {\bf 63}  013607 (2000); J. Wachter, R. Walser, J. Cooper, and M. Holland Phys. Rev. A  {\bf 64}  053612 (2001)

\bibitem{Zaremba} B. Jackson and E. Zaremba Phys. Rev. Lett. {\bf 88} 180402 (2002)


\bibitem{Krauth} W. Krauth Phys. Rev. Lett. {\bf 77} 3695 (1996)

\bibitem{Ceperley} D. M. Ceperley, Rev. Mod. Phys. {\bf 71} S438 (1999)

\bibitem{Walls} M. J. Steel, M. K. Olsen, L. I. Plimak, P.D. Drummond, S.M. Tan, M. J. Collett, D.F. Walls, and R. Graham Phys. Rev. A {\bf 58} 4824 (1998);

\bibitem{Drummond} P.D. Drummond and J. F. Corney Phys. Rev. A {\bf 60} R2661 (1999)


\bibitem{Carusotto} I. Carusotto, Y. Castin, and J. Dalibard Phys. Rev. A  {\bf 63}  023606 (2001)

\bibitem{TDHF} V. Chernyak and S. Mukamel J. Chem. Phys. {\bf 104}, 444 (1996);
 S. Tretiak, V. Chernyak, and S. Mukamel J. Am. Chem. Soc. {\bf 119} 11408 (1997);
 S. Tretiak and S. Mukamel, Chem. Rev. {\bf 102}, 3171 (2002).

\bibitem{QHE} {\em The Quantum Hall Effect}, edited  by R.~E. Prange and
S.~M. Girvin (Springer-Verlag,  N.Y. 1987).

\bibitem{Zhang} S.~C. Zhang, T.~H. Hansson and S. Kivelson, Phys. Rev. Lett. {\bf 62}, 82 (1989).

\bibitem{Lopez} A. Lopez and E. Fradkin, Phys. Rev.  {\bf B 44}, 5246 (1991).


\bibitem{Meng} H.~F. Meng, Phys. Rev.  {\bf B 49}, 1205 (1994).

\bibitem{Laughlin} R.~B. Laughlin, Phys. Rev. Lett. {\bf 50}, 1395 (1983).

\bibitem{Jain} J.~K. Jain, Phys. Rev. Lett.  {\bf 63}, 199 (1989).


\bibitem{HLR} B.~I. Halperin, P.~A. Lee and N. Read, Phys. Rev.  {\bf B 47}, 7312 (1993).

\bibitem{Birman} N.~A. Zimbovskaya and J.~L. Birman,   Phys. Rev. {\bf B 60}, 16762 (1999).

\bibitem{Birman1} N.~A. Zimbovskaya and J.~L. Birman,   Int. J. Mod. Phys. {\bf B 13}, 859 (1999).


\bibitem{ExRb}  D. S. Jin, J. R. Ensher, M. R. Matthews, C. E. Wieman and E.
A. Cornell, Phys. Rev. Lett. {\bf 77}, 420 (1996)

\bibitem{ExNa}  M.-O. Mewes, M. R. Andrews, N. J. van Druten, D. M. Kurn, D.
S. Durfee and W. Ketterle, Phys. Rev. Lett. {\bf 77}, 416 (1996)


\bibitem{Arnoldi} F. Chatelin, {\it Eigenvalues of Matrices} (Wiley, New
York, 1993); V. Chernyak, M. F. Schulz, and S. Mukamel, J. Chem.
Phys. {\bf 113} 36 (2000)

\bibitem{Arfken} G. B. Arfken and H.J. Weber, {\it Mathematical methods for
physicists} 4th Ed. (Academic Press, San Diego, 1995)

\bibitem{Golub} G. H. Golub and C. F. Van Loan, {\it Matrix Computations}
(The Johns Hopkins University Press, Baltimore, 1983)



\bibitem{burnett}  N.P. Proukakis and K. Burnett, J. Res. Natl. Inst. Stand.
Technol. {\bf 101}, 457 (1996)




\bibitem{perelomov} A. M. Perelomov, {\em Generalized Coherent States and Their Applications} (Springer-Verlag, New York, 1986)


\bibitem{Chernyak} V. Chernyak and S. Mukamel, J. Chem. Phys. {\bf 111}, 4383
(1999).

\bibitem{ChernyakChoi} V. Chernyak, S. Choi and S. Mukamel, Phys. Rev. {\bf A} {\bf 67}, 053604 (2003).





\end{thebibliography}
\end{document}